\pdfoutput=1
\documentclass{article}

\usepackage{graphicx} 

\usepackage{subcaption}
\usepackage{natbib}
\setcitestyle{authoryear,open={(},close={)}}	

\usepackage{amsthm}
\usepackage[a4paper]{geometry}

\usepackage[figuresright]{rotating}

\usepackage[utf8]{inputenc} 
\usepackage{bm}
\usepackage{amsmath}

\usepackage{booktabs}
\usepackage{multirow}
\usepackage{xcolor}
\usepackage[english]{babel}
\usepackage{csquotes}
\usepackage{MnSymbol}
\usepackage{pifont}

\definecolor{altbrown}{rgb}{0.65, 0.16, 0.16}

\makeatletter
\def\blfootnote{\xdef\@thefnmark{}\@footnotetext}
\makeatother

\newcommand{\bz}[0]{\mathbf{Z}}
\newcommand{\bn}[0]{\bm{\beta_0}}
\newcommand{\bh}[0]{\bm{\hat{\beta}}}
\newcommand{\bb}[0]{\bm{\beta}}
\newcommand{\xb}[0]{\bm{X}}
\newcommand{\kh}[0]{\hat{\kappa}}
\newcommand{\ka}[0]{\kappa}
\newcommand{\an}[0]{\bm{\alpha_0}}
\newcommand{\ah}[0]{\bm{\hat{\alpha}}}
\newcommand{\ab}[0]{\bm{\alpha}}

\newcommand{\ubark}[0]{\bm{\bar{U}}_{n, \kappa}}

\newcommand{\bsig}[0]{\bm{\Sigma}}
\newcommand{\ubm}[0]{\bm{U}}

\newcommand{\balpha}{\boldsymbol{\alpha}}
\newcommand{\bv}{\boldsymbol{v}}
\newcommand{\be}{\boldsymbol{e}}
\newcommand{\bs}{\boldsymbol{s}}
\newcommand{\bV}{\boldsymbol{V}}
\newcommand{\bU}{\boldsymbol{U}}
\newcommand{\bX}{\boldsymbol{X}}
\newcommand{\bE}{\boldsymbol{E}}
\newcommand{\bZ}{\boldsymbol{Z}}
\newcommand{\bS}{\boldsymbol{S}}
\newcommand{\bO}{\boldsymbol{O}}
\newcommand{\bQ}{\boldsymbol{Q}}
\newcommand{\bA}{\boldsymbol{A}}
\newcommand{\bD}{\boldsymbol{D}}

\newcommand{\bI}{\boldsymbol{I}}
\newcommand{\bcalB}{\boldsymbol{{\cal B}}}
\newcommand{\bmu}{\boldsymbol{\mu}}
\newcommand{\ba}{\boldsymbol{a}}
\newcommand{\bOmega}{\boldsymbol{\Omega}}

\begin{document}

\title{Causal Proportional Hazards Estimation with a Binary Instrumental Variable}

\author{
Behzad Kianian$^{1, +}$, Jung In Kim$^{2, +}$, Jason P. Fine$^{2}$, \\ Limin Peng$^{1, *}$ \\ 
\\
	$^{1}$Department of Biostatistics and Bioinformatics, Emory University, \\ 
	Atlanta, U.S.A. \\
	$^{2}$Department of Biostatistics, University of North Carolina at Chapel Hill, \\ 
	Chapel Hill, U.S.A. \\
}

\date{}


\maketitle

\blfootnote{$^{+}$The first two authors have equal contributions to this work}
\blfootnote{$^{*}$Corresponding author: lpeng@emory.edu}

\begin{abstract}
	Instrumental variables (IV) are a useful tool for estimating causal effects in the presence of unmeasured confounding. IV methods are well developed for uncensored outcomes, particularly for structural linear equation models, where simple two-stage estimation schemes are available. The extension of these methods to survival settings is challenging, partly because of the nonlinearity of the popular survival regression models and partly because of the complications associated with right censoring or other survival features. We develop a simple causal hazard ratio estimator in a proportional hazards model with right censored data. The method exploits a special characterization of IV which enables the use of an intuitive inverse weighting scheme that is generally applicable to more complex survival settings with left truncation, competing risks, or recurrent events. We rigorously establish the asymptotic properties of the estimators, and provide plug-in variance estimators. The proposed method can be implemented in standard software, and is evaluated through extensive simulation studies. We apply the proposed IV method to a data set from the Prostate, Lung, Colorectal and Ovarian cancer screening trial to delineate the causal effect of flexible sigmoidoscopy screening on colorectal cancer survival which may be confounded by informative noncompliance with the assigned screening regimen.
\end{abstract}

\emph{keywords}: Causal treatment effect; Cox proportional hazards model; Instrumental variable (IV); Noncompliance.

\section{Introduction}
\label{sec:intro}

Research studies are often fundamentally interested in understanding the causal effect of  a treatment or exposure on an outcome of interest \citep{Holland1986}. In observational studies, unmeasured confounding is a major obstacle to estimating the causal effect of a nonrandomized exposure on disease etiology. Such a challenge also arises in well-designed randomized clinical trials. 
When there are issues of non-compliance in the treatment arms, the treatment decision may be based on latent (unobserved) factors that strongly correlate with clinical outcomes. This would result in bias from unmeasured confounding and hence  complicate the task of estimating the ``efficacy'' of the treatment. 

Instrumental variables (IVs) offer a useful tool for estimating causal treatment or exposure effects in these settings \citep{Angrist1995, Angrist1996,  Loeys2003, Li2015, Li2016, Mackenzie2016}. Informally, IVs have the characteristics of being independent of unmeasured confounders, being related to the treatment, and only being related to the outcome through the treatment \citep{Baiocchi2014}.  In observational studies, there are a variety of potential sources for instruments that can aid in the estimation of causal effects, either of treatment or exposure \citep{Baiocchi2014}. In randomized clinical trials with non-compliance, the treatment assignment mechanism can serve as an instrumental variable.

The motivating example of this work is the Prostate, Lung, Colorectal and Ovarian (PLCO) Cancer screening trial, which is a multi-center randomized trial designed to evaluate the effectiveness of the screening with flexible sigmoidoscopy compared versus usual care. In this study, $77,449$ subjects were randomly assigned to the intervention group,
but only 85\% complied with the assigned  sigmoidoscopy protocol. Such non-compliance may be outcome-related. For example, relatively healthy individuals may be more likely to skip the screening.  In the presence of unmeasured confounding, neither intent-to-treat (ITT) analysis nor ``as-treated'' analysis would be adequate for assessing the causal benefit of the treatment (i.e. flexible sigmoidoscopy screening). A possible remedy is an IV analysis that properly adjusts for the selection bias induced by subjects' post-randomization care selection. The assigned treatment in a randomized trial serves a natural instrumental variable which may be utilized in this analysis.


IV methodology has primarily focused on linear models and continuous outcomes in contexts without censoring. Recently, research on IV methodology for time-to-event data with right censoring has grown rapidly. For example, \cite{Baker1998} developed an IV method for randomized trials with all-or-none compliance and discrete-time survival data by estimating the hazards for compliers under treatment and control in a manner analogous to two-stages least squares (TSLS) methods for linear models. \cite{Baiocchi2014} gave a brief summary of this method in their review of IV methods.  Building on \cite{Baker1998}'s work, \cite{Nie2011} developed an estimation method with improved efficiency.

\cite{Robins1991} considered a structural accelerated failure time model and developed estimators for the causal treatment effect in the context with non-compliance and only administrative censoring. \cite{Joffe2001} provided a detailed discussion of this general approach. Imposing parametric distributional assumptions, \cite{Li2015} developed a Bayesian approach for IV analysis with censored time-to-event outcome under a two-stage linear model. \cite{LiFine2015} and \cite{Tchetgen2015} developed IV based methods under additive hazards modeling of time-to-event data. 
Specifically, \cite{LiFine2015} proposed a two-stage, consistent estimation procedure for a causal treatment effect by substituting the predicted treatment from the observed covariates and IV.  \cite{Tchetgen2015} developed a ``control function'' approach, where the residual from a model with the treatment as the outcome was added to the additive hazards regression model. 
More recently, \cite{Martinussen2016} studied structural cumulative survival models with time-varying exposures in a specification that was related to the additive hazards model.

In time-to-event analysis, proportional hazards model is the most popular formulation for the effects of treatment and covariates.
There have been several IV approaches developed under the proportional hazards modeling. For example, for the special case of all or none noncompliance without covariates, \cite{Loeys2003} proposed an estimate for the complier proportional hazards effect of treatment by deriving a properly imputed partial likelihood that recovered the unobserved information on the treatable subgroup in the control arm.  Also working in the noncompliance setting, \cite{Cuzick2007} constructed a Mantel-Haenszel-type estimator for the case without covariates and a partial-likelihood based estimator when covariates were present and independent of compliance types. A full likelihood based approach was explored for situations where covariates were correlated with compliance type.  \cite{Li2016} further proposed an EM algorithm for the full likelihood based estimation.
\cite{Yu2015} tackled the problem of estimating causal estimands including the complier average causal effect, complier survival probability, and complier quantile causal effect under the semiparametric transformation model. They adapted the nonparametric likelihood estimation technique of \cite{Zeng2007}, and provided an EM algorithm for implementing the proposed estimation  as well as  theoretical justifications.
While the likelihood-based strategies accommodate both censoring and covariates in the estimation of the causal treatment effect with censored time-to-event data, the resulting estimation and inference procedures are generally very complicated. The computational complexity and stability may become unbearable when the sample size is large, such as in the PLCO Cancer screening trial. Furthermore, they require specifying causal models for all latent compliance classes, not just that of interest, which may impair the  robustness of these methods to potential model misspecification.


In this work, we develop a  new IV approach to estimating a causal treatment effect under the proportional hazards modeling of time-to-event outcome subject to independent right censoring. The causal estimand  is defined within the latent subgroup of compliers as in  most existing work on causal proportional hazards regression. However, our method does not need to impose regression models for the latent compliance classes other than the complier subgroup. Our key strategy is to adapt the seminal work of \cite{Abadie2003}  which provides a simple link between the unconditional moment of the observed data and the conditional moment given the latent complier group.  \cite{Abadie2003} developed a simple weighting strategy which is easily applied to estimating equations which are sums of independent terms.
However, an analogous application to the proportional hazards regression is not straightforward. This is because the partial likelihood does not yield an estimating function of the simple form as a sum of independent terms, as with the least squares criterion  for linear regression. To circumvent this difficulty, we take carefully designed steps to incorporate the weighting idea of \cite{Abadie2003} through the asymptotic influence functions of the partial likelihood score equation. We establish the large sample properties of the resulting parameter estimators, including consistency and asymptotic normality. To attain stable and fast computation, we make further efforts to devise computational algorithms to obtain the proposed parameter estimators. The calculations can be easily implemented via existing software for weighted proportional hazards regression. {We also illustrate that the proposed weighting scheme is generally applicable to more complex survival settings, for example, in the presence of left truncation, competing risks, or recurrent events. Such a broad applicability appears  lacking in existing IV approaches for proportional hazards models.}

In Section 2, we  first introduce the potential outcomes framework including the latent compliance groups, the IV assumptions, and the set-up of causal proportional hazards regression.  We next describe the proposed estimation procedure {with randomly censored data}, discuss computational considerations, and present a modification of the proposed method which has improved computational features. {Adaptations to settings with left truncation, competing risks, or recurrent events are also discussed}. We  rigorously present the consistency and asymptotic normality of the estimators. The results include a closed form for the asymptotic variance of the estimator and a consistent plug-in variance estimator. Bootstrap variance estimates are also provided.  The results from extensive simulations are reported Section 3 and demonstrate that the methods perform well with realistic sample sizes. In Section 4, we apply our methods to the data from the PLCO Cancer screening trial. Some remarks conclude in Section 5.

\section{Weighted Partial Likelihood Estimation for Causal Proportional Hazards Models} \label{sec:method}

\subsection{Potential Outcomes Framework}

We introduce the potential outcomes framework and notation commonly employed in the causal inference literature. Consider potential survival times $T_1$ and $T_0$ based on receiving ($D=1$) and not receiving the treatment ($D=0$), respectively. Define $V$ as a binary IV, and define the potential treatment $D_v$ such that $D_1$ denotes the treatment received when $V=1$ and $D_0$ denotes the treatment received when $V=0$. Following the terminology of \cite{Abadie2003}, subjects can be classified into 4 latent compliance groups based on the potential treatment indicators: \emph{compliers} ($D_1 > D_0$), \emph{always-takers} ($D_1 = D_0 = 1$), \emph{never-takers} ($D_1 = D_0 = 0$), and \emph{defiers} ($D_1 < D_0$). In the PLCO Cancer screening trial,  compliers would be the individuals who were assigned to the intervention group and also took the flexible sigmoidoscopy screening. Always-takers (or never-takers) are defined as always (or never) taking the flexible sigmoidoscopy screening. Defiers are individuals who would take the the flexible sigmoidoscopy screening if assigned to the usual care group but not if assigned to the intervention group. Since $D_1$ and $D_0$ cannot be observed at the same time, we are not able determine the latent compliance group membership of any individual based on the observed data alone.

Define the potential outcome for each subject as $T_{vd}$, which represents the survival time $T$ if $V=v$ and $D=d$. Let $\xb$ represent the covariate vector. We re-state several key assumptions from \cite{Abadie2003} about the IV, $V$:
\newtheorem*{assumption*}{Assumptions (A1)-(A4)}
\begin{assumption*} Let $T_{vd}, \xb, V, D, D_v$ be defined as above. \label{as:A1A4}
	\begin{enumerate}
		\item[(A1)] \emph{Independence of the instrument:} \[(T_{00}, T_{01}, T_{10}, T_{11}, D_{0}, D_{1}) \perp V | \xb \]
		\item[(A2)] \emph{Exclusion of the instrument:} \(P(T_{1d} = T_{0d}|\xb) = 1\) for \(d = 0, 1\).
		\item[(A3)] \emph{First stage:} \( 0 < P(V=1|\xb) < 1 \) and \( P(D_{1} = 1 | \xb) > P(D_{0}=1|\xb) \)
		\item[(A4)] \emph{Monotonicity:} \(P(D_{1} \geq D_{0} | \xb) = 1 \)
	\end{enumerate}
\end{assumption*}

Assumption (A1) says that the instrument $V$ is as good as random conditional on the covariates $\xb$, or equivalently, that $V$ is independent of unmeasured confounders conditional on $\xb$. Assumption (A2) says that the instrument $V$ only influences the outcome $T$ through its effect on the treatment $D$. Assumption (A3) states that every subject has some chance of receiving the instrument $V$, conditional on the covariate $\xb$, and that conditional on $\xb$, $V$ has an effect on the treatment received. Finally, assumption (A4) says that with probability one, defiers do not exist.



\subsection{Model Formulation}

Our focus is to estimate and make inferences about the treatment effect for the latent group of compliers. Specifically, we adopt Cox's proportional hazards regression model to formulate the effects of  treatment and covariates for compliers:
\begin{equation} \label{coxmodel}
h(t; D, \xb) = h_{0}(t) \exp\{\beta_d D + \bm{\beta}_x^{T} \xb\},
\end{equation} where $h(t; D, \xb)$ is the hazard function for compliers defined as
$$
h(t; D, \xb) = \lim_{\Delta t \rightarrow 0} \frac{\Pr(t \leq T \leq t + \Delta t| T \geq t, D_1 > D_0, D, \xb)}{\Delta t},
$$
and $h_0(t)$ is an unspecified baseline hazard at time $t$. 
In model \eqref{coxmodel}, $\beta_d$  is the causal estimand of the primary interest, which can be interpreted as the causal treatment effect for compliers after adjusting for the covariate effects captured by $ \bm{\beta}_x$ \citep{Abadie2003}. Such a quantity has frequently been of interest in literature 
\citep[for example]{Loeys2003, Cuzick2007, Yu2015}.
It is worth emphasizing that the proportional hazards model \eqref{coxmodel} is only assumed for compliers. In contrast, likelihood-based approaches
\citep[for example]{Cuzick2007, Yu2015, Li2016} typically require distributional modeling for the other compliance subgroups (e.g. {\em always takers}, {\em never-takers}) and may be biased under misspecification of those models. 

\subsection{Estimation}

In practice, $T$ is often subject to right censoring by $C$; thus we observe $W = min(T,C)$ and $\delta=I(T \leq C)$ instead of $T$. We adopt the standard random censoring assumptions that $C$ is independent of $T$ conditional on $(V, D, \xb)$. We further assume that $C$ is independent of $V$ given $\xb$.
Defined ${\bO} = (W, \delta, D, \xb, V)$. The observed data consist of $n$ independently identically distributed (i.i.d.) replicates of ${\bO}$, denoted by  $\{\bO_i\}_{i=1}^{n} = \{(W_i, \delta_i, D_i, \xb_i, V_i)\}_{i=1}^n$. Define $Y_i(t) = I(W_i \geq t)$ and $N_i(t) = I(W_i \leq t, \delta_i=1)$, which represent the at-risk process and the observed event counting process for subject $i$ respectively.
We also assume that there are no ties (i.e. $dN_i(t)\leq 1$). In the sequel,  we use the subscript $i$ to differentiate population quantities and their sample analogues throughout the paper.

Let $\bn=(\beta_d, \bm{\beta}_x)$ and $\bZ=(D, \xb)$. When all subjects are known to be compliers, the estimation of $\bn$ can proceed through standard Cox regression analysis  \citep{Andersen1982}. This is because, in this case, the hazard function for the whole study population,
$
\lambda(t|\bZ)\equiv\lim_{\Delta t \rightarrow 0} {\Pr(t \leq T \leq t + \Delta t| T \geq t, D, \xb)}/{\Delta t},
$
equals that for the latent complier subgroup, $\exp(\bb^T\bZ)$ $h_0(t)$. Then $M(t)\equiv N(t)-\int_0^t Y(s)\exp(\bb_0^T\bZ)h_0(s)ds$ is a martingale, and thus
a consistent estimator  of $\bb_0$ 
can be obtained as the solution of the partial likelihood score equation,
\begin{equation} \label{Ubasic1}
\bm{U}_n(\bb) =  \frac{1}{\sqrt{n}}\sum_{i=1}^n \int_0^{\infty} \left\{ \bz_i - \frac{  \bm{S}_n^{(1)}(\bb, s) } { S_n^{(0)}(\bb, s) }  \right\}  \,dN_i(s),
\end{equation}
where $\bm{S}_n^{(j)}(\bb, s) = \sum_{l=1}^n Y_{l}(s) \bz_{l}^{\otimes j} e^{\bb^{T}\bz_{l}}$ for $j = 0, 1, 2$. Here and in the sequel, for a vector $\bv$, $\bv^{\otimes 0}=1$, $\bv^{\otimes 1}=\bv$, and $\bv^{\otimes 2}=\bv\bv^T$.


Next we consider the more realistic case where the study population consists of both compliers and non-compliers.
In this case,  $\lambda(t|\bZ)$ generally deviates from the hazard function assumed for the complier group, $h_0(t)\exp(\bb_0^T\bZ)$. As a result,  $M(t)$ is no longer a martingale for the whole study population, and
equation \eqref{Ubasic1} would fail to provide a valid estimate for $\bb_0$.

To construct an appropriate estimating equation for $\bn$, we utilize the fact that $M(t)$ remains a martingale for the complier group. Using this fact, we can show 
that
$\bmu_c(\bb_0)=0$ under model \eqref{coxmodel}, where
$\bm{s}_{c}^{(j)}(\bb, s) = E(Y(s) \bz^{\otimes j} e^{\bb^{T}\bz} | D_1 > D_0)$ ($j=0, 1, 2$) and $$
\bmu_c(\bb)=E\left[\int_0^\infty \left\{\bZ-\frac{ \bm{s}_{c}^{(1)}(\bb, s)} { s_{c}^{(0)}(\bb, s)} \right\}dM(s)\bigg | D_1>D_0\right].
$$


However,  $\bmu_c(\bb)$ cannot be directly used to estimate $\bb_0$ because the latent complier group, $\{D_1>D_0\}$, is not observed. To tackle this difficulty,
we adopt the strategy of  \cite{Abadie2003}, which  established a simple link between the unconditional moment of the observed data and the conditional moment of the data within the complier group. A simple weighting approach may be employed to identify the regression parameters associated with the complier group. 
More specifically, let $g(\cdot)$ be a measurable real function of $(T, D,\xb, C)$ such that $E|g(T, D, \xb, C)| < \infty$. Under assumptions (A1)--(A4) and given $C$ is independent of $V$ given $\xb$, Theorem 3.1 of \cite{Abadie2003} implies that
\begin{equation}
\label{Abadie}
E\{g(T, D,\xb, C)|D_{1} > D_{0}\}= \frac{1}{\Pr(D_{1} > D_{0})} E\{\kappa\cdot g(T, D, \xb, C)\}
\end{equation}
where
\begin{equation} \label{kappa}
\kappa = 1 - \frac{D(1-V)}{\Pr(V=0|\xb)} - \frac{(1-D)V}{\Pr(V=1|\xb)}.
\end{equation}
This result suggests that a weighting scheme involving $\kappa$ can lead to the identification of moment-type statistics for compliers.
One should recognize that $\kappa$ can take both positive and negative values. This differs from standard weighting procedures based on probability weighting, where the weights are always positive as a result of probabilities being nonnegative. This creates nonstandard computational challenges, which are discussed further below.

Using \eqref{Abadie}, we obtain the following key results for deriving an estimating equation for $\bb_0$:
$$
\bmu_c(\bb)={1\over Pr(D_1>D_0)}E\left[\kappa\int_0^\infty \left\{\bZ-\frac{ \bm{s}_{c}^{(1)}(\bb, s)} { s_{c}^{(0)}(\bb, s)} \right\}dM(s)\right],
$$
where
$$
\bm{s}_{c}^{(j)}(\bb, s) =  \frac{E(\kappa Y(s) \bz^{\otimes j} e^{\bb^{T}\bz})}{\Pr(D_1 > D_0)},\ \ j=0, 1, 2.
$$

Suppose $\kappa_i$ is known for each subject $i$. One may construct a weighted estimating equation  for $\bb_0$, $\bU_{n, \kappa}(\bb)=0$, where
$$
\ubm_{n, \kappa}(\bb) =  \frac{1}{\sqrt{n}}\sum_{i=1}^n \int_0^{\infty} \kappa_i \left( \bz_i  -  \left\{ \frac{ \bm{S}_{n,\kappa}^{(1)}(\bb,s) } {  S_{n,\kappa}^{(0)}(\bb,s) }  \right\} \right)\,dN_i(s)
$$
with $ \bm{S}_{n, \ka}^{(j)}(\bb, s) =  \sum_{l=1}^n \ka_l Y_{l}(s) \bz_{l}^{\otimes j} e^{\bb^{T}\bz_{l}}$. Note that $\ubm_{n, \kappa}(\bb)$ remains the same if $dN_i(s)$ is replaced by $dM_i(s)$, and hence $\ubm_{n, \kappa}(\bb)$ is proportional to an empirical counterpart of $\bmu_c(\bb)$. This justifies the use of $\ubm_{n, \kappa}(\bb)$ for constructing the estimating equation for $\bb_0$.


In general, $\kappa_i$'s is known a priori, 
for example, with external information. In practice, we propose to estimate $\kappa_i$ by imposing additional modeling assumptions for $\Pr(V=1|\bX)$. Specially, we may assume a logistic regression model for $V$:
\begin{equation}
P(V=1|\xb) \equiv \psi(\bm{\alpha}_0, \xb)  = \frac{\exp(\alpha_{01} + \bm{\alpha}_{02}^{T}\xb)}{1 + \exp(\alpha_{01} + \bm{\alpha}_{02}^{T}\xb)},
\label{LR_model}
\end{equation}
with $\balpha_0=(\alpha_{01}, \balpha_{02}^T)^T$.
Let  $\hat{\bm{\alpha}}$ be the maximum likelihood estimator of $\bm{\alpha}_0$ \citep{Gourieroux1981, Agresti2013} and define
\begin{equation} \label{kappahat}
\kh_i = 1 - \frac{D_i(1-V_i)}{1 - \psi(\hat{\bm{\alpha}},\xb_i)} - \frac{(1-D_i)V_i}{\psi(\hat{\bm{\alpha}},\xb_i)}.
\end{equation}
Replacing the $\kappa_i$ in $\bU_{n, \kappa}(\bb)$ by $\hat\kappa_i$ leads to the proposed estimating equation:
\begin{equation}\label{proposed_ee}
\bU_{n, \hat\kappa}(\bb)=0,
\end{equation}
where
\begin{equation} \label{eefinal}
\ubm_{n, \hat{\kappa}}(\bb) =  \frac{1}{\sqrt{n}}\sum_{i=1}^n \int_0^{\infty} \kh_i \left( \bz_i  -  \left\{ \frac{ \bm{S}_{n,\kh}^{(1)}(\bb,s) } {  S_{n,\kh}^{(0)}(\bb,s) }  \right\} \right)\,dN_i(s).
\end{equation}
Denote the solution to equation \eqref{proposed_ee} by $\hat\bb$.
The detailed computational algorithm for obtaining $\hat\bb$, and the related algorithmic issues and remedies are discussed in the next subsection.

\subsection{The Computational Algorithm}

The form of the proposed estimation equation \eqref{proposed_ee} closely resembles the estimating equation for a weighted Cox proportional hazards regression. However, an important distinction is that  $\kh_i$'s in   \eqref{proposed_ee} can take negative values. As a result, $\ubm_{n,\kh}(\bb)$ can have a highly irregular surface with multiple zero-crossings. To address this complication, we propose to locate $\hat\bb$ through finding the maximizer of a properly designed objective function.
Specifically, instead of directly solving $\ubm_{n,\kh}(\bb) = 0$, we propose to obtain $\bh$ as the maximizer of the following objective function
\begin{equation} \label{objfun}
\bar{C}_{n, \kh}(\bb) = \frac{1}{n} \sum_{i=1}^{n} \kh_i \delta_i \left[ \bb^{T}\bz_i  - \log\{ \tilde{S}_{n, \kh}^{(0)}(\bb, W_i)  \} \right],
\end{equation}
where $\tilde{S}_{n, \kh}^{(0)}(\bb, t)=\max(S_{n, \kh}^{(0)}(\bb, t), \nu)$ and $\nu$ is a pre-specified small positive value. The justification for doing so is that $\partial\bar{C}_{n, \kh}(\bb)/\partial\bb^T$ would be nearly the same as $n^{-1/2}\bU_{n, \hat\kappa}(\bb)$ because $\nu$ can be arbitrarily small.  Truncating  $S_{n, \kh}^{(0)}(\bb, t)$ below by $\nu$ ensures the positiveness of the resulting quantity.  In theory, the asymptotic limit of $S_{n, \kh}^{(0)}(\bb, t)$ is strictly positive under mild regularity conditions. Therefore, such a truncation should have  negligible impact on the finite-sample performance of $\hat\bb$ when $n$ is reasonably large. In our numerical studies, we choose $\nu=10^{-4}$.

The procedure for obtaining $\hat\bb$ is as follows.
\begin{enumerate}
	\item Fit the logistic regression model \eqref{LR_model} to $\{(V_i, \xb_i)\}_{i=1}^n$ and obtain $\hat\balpha$.
	\item Calculate $\hat\kappa_i$ using formula \eqref{kappahat}.
	\item Find the maximizer of  the objective function $\bar C_{n, \hat\kappa}(\bb)$ in \eqref{objfun} by an optimization routine, such as {\em optim()} function in R \citep{rbase}.
\end{enumerate}

\subsection{A Modified Weighting Scheme}

In principle,  the objective function $\bar C_{n, \hat\kappa}(\bb)$ approaches a limit that is concave, and standard optimization routines are expected to work well when the sample size is large. However, the presence of negative weights $\kappa_i$'s  can sometimes lead to a highly irregular surface for $\bar{C}_{n, \kh}(\bb)$ and $U_{n,\kh}(\bb)$ (see figures in Appendix C) 
and result in numerical instability for estimating $\bb_0$. To address this problem, we propose a modified weighting scheme, which can avoid negative weights and allow us to obtain $\hat\bb$ through standard computational routines for the weighted proportional hazards regression, such as the {\em coxph()} function in R \citep{rsurvival}.


Let $\bU=( W, \delta, D, \xb)$. We define a modified weight by projecting the original weight $\ka$ as follows:
\begin{equation} \label{modkappa}
\kappa_v = E(\kappa|\bU) = 1 - \frac{D(1 - v_0(\bU))}{P(V=0|\xb)} - \frac{(1 - D)v_0(\bU)}{P(V=1|\xb)},
\end{equation} where $v_0(\bU) = E(V|\bU) = P(V=1|W, \delta, D, \xb)$.  Adapting the arguments of \cite{Abadie2002}, we can show that $\kappa_v= P(D_1 > D_0|\bU)$  and $\kappa_v$ can play the same role as $\kappa$ in equation \eqref{Abadie} (see Appendix A). This result indicates that $\kappa_v$ is a probability; thus it is always non-negative and can be regarded as a proper weight. Adopting the weighting scheme by $\kappa_v$ can avoid the potential numerical issues present with $\kappa$. 
We propose to estimate $\ka_v$ as follows.
\begin{enumerate}
	\item Stratify the data by the censoring and treatment status: $\{(\delta=c, D=d)\}$, $c = 0,1$, $d=0,1$.
	\item Within each stratum,  fit a nonparametric or parametric regression model for $V$ given covariates $(W, \xb)$. This will provide an estimate for $v_0(\bU)$, denoted by $\hat v(\bU)$.
	\item Calculate the estimated $\kappa_v$ as
	$$
	\hat\kappa_v=1 - \frac{D(1 - \hat v(\bU))}{1-\psi(\hat\balpha, \bX_i)} - \frac{(1 - D)\hat v(\bU)}{\psi(\hat\balpha, \bX_i)}.
	$$
\end{enumerate}

In Step 2 above,  we may consider non-parametric power series  (NPPS) regression or logistic regression for $V$ given $(W, \bX)$.  Based on our extensive numerical experiences (including those reported and not reported in Section \ref{sec:simstudy}), a second-order logistic regression model with the interaction between $W$ and $\xb$ works well compared to approaches that estimate $\kh_v$ from NPPS, or the first-order logistic regression.
Note that, with finite sample sizes, the resulting estimator $\hat{\kappa}_v$ may be negative or greater than $1$. To circumvent the undesirable numerical properties associated with negative weights, we  propose a slightly different modified weight, $\hat{\kappa}_{v,tr}$, that truncates $\hat{\kappa}_v$ such that its value lies strictly in  an interval ${\cal I}\subset(0,1)$, say $[0.01, 0.99]$. Since the true weight $\kappa_v$ is between $0$ and $1$ and we can let ${\cal I}$ be arbitrarily close to $(0, 1)$,  there should be negligible asymptotic bias induced by such a truncation. Using $\kh_{v, tr}$ in place of $\kappa$ in \eqref{proposed_ee},  we can easily obtain $\bh$ through the R function, {\em coxph()}, with the weight argument properly specified. In Section \ref{sec:simstudy}, we thoroughly examine the performance of $\hat\bb$ with different choices of weight.

\subsection{Generalizations to complex survival settings}
{
Survival data are often subject to complications other than random right censoring, for example, left truncation, competing risks, and recurrent events. The proposed weighting scheme can be readily adapted to accommodate these additional data complexities.

{\bf Left truncation:}
Suppose the survival time $T$ is subject to left truncation by $L$. We observe $\tilde O \equiv (\tilde W, \tilde\delta, \tilde L, \tilde D, \tilde X, \tilde V)$, where $\tilde O$ follows the conditional distribution of $(W, \delta, L, D, X, V)$ given $L< W$.  Let $\tilde O_i\equiv (\tilde W_i,  \tilde\delta_i, \tilde L_i, \tilde D_i, \tilde X_i, \tilde V_i)$ be the sample analogue of $\tilde O$.
Assume that $(L, C)$ is independent of $T$ given $(V, D, X)$, and $(L, C)$ is independent of $V$ given $X$.

\sloppy Define $\tilde N(t)= I(\tilde L<\tilde W\leq t, \delta=1)$, $\tilde Y(t)= I(\tilde W\geq t> \tilde L)$, and $\tilde M(t)\equiv \tilde N(t)- \allowbreak \int_0^t \tilde Y(s)\exp(\bb_0^T \bZ)h_0(s)ds$. The partial likelihood score equation under left truncation \citep{Andersen2012} suggests that $\tilde\bmu_c(\bb_0)=0$, where
$$
\tilde\bmu_c(\bb)=E\left[\int_0^\infty \left\{\tilde\bZ-\frac{ \tilde{\bm{s}}_{c}^{(1)}(\bb, s)} { \tilde s_{c}^{(0)}(\bb, s)} \right\}d\tilde M(s)\bigg | D_1>D_0, L<W\right],
$$
where $\tilde\bs_{c}^{(j)}(\bb, s) = E(\tilde Y(s) \tilde\bz^{\otimes j} e^{\bb^{T}\tilde\bz} | D_1 > D_0, L<W)$ ($j=0, 1, 2$).  Applying the same technique shown in \eqref{Abadie}, one can establish that
\begin{equation}
\label{truncation}
\tilde\bmu_c(\bb)=\frac{1}{\Pr(D_1>D_0|L<W)}E\left[\tilde\kappa\int_0^\infty \left\{\tilde\bZ-\frac{ \tilde\bs_{c}^{(1)}(\bb, s)} { \tilde s_{c}^{(0)}(\bb, s)} \right\}d\tilde M(s)\bigg |L<W\right],
\end{equation}
where
$$
\tilde\kappa=1-\frac{\tilde D(1-\tilde V)}{\Pr(\tilde V=0|\tilde\bX, L<W)}-\frac{(1-\tilde D)\tilde V}{\Pr(\tilde V=1|\tilde \bX, L<W)}
$$
and
$$
\bm{s}_{c}^{(j)}(\bb, s) =  \frac{E(\kappa_l  \tilde Y(s) {\tilde\bz}^{\otimes j} e^{\bb^{T}{\tilde\bz}}|L<W)}{\Pr(D_1 > D_0, L<W)},\ \ j=0, 1, 2.
$$

The result in \eqref{truncation} suggests a simple adaptation of the proposed method to the case with random left truncation, where the main modification is to  replace $\bZ_i$, $Y_i(t)$, $N_i(t)$ in $U_{n, \kappa}(\bb)$ by $\tilde\bZ_i$, $\tilde Y_i(t)$, $\tilde N_i(t)$ respectively. The weights $\hat\kappa$ or $\hat\kappa_v$ can be calculated in the same way as in Section 2.3 and 2.5 based on $\tilde D_i, \tilde V_i, \tilde\bX_i, \tilde W_i$ observed under left truncation.

{\bf Competing risks:} Consider a typical competing risks setting with $K$ types of competing failures. Let $T=\min(T_1,\ldots, T_K)$, where $T_k$ denotes the latent event time to failure type $k$ ($k=1,\ldots, K$). Let $C$ denote time to random censoring for $T$, which satisfies the same censoring assumptions  stated in Section 2.3. Let $W=\min( T, C)$ and define $\eta$ as $0$ if $T>C$ and  the type of failure otherwise.  We observe $(T, \eta, V, D, \bX)$.

When the interest lies in the minimal event time $T$, one can simply  apply the procedures in Section 2.3-2.5 to the observed data on $(T, I(\eta\neq 0), V, D, \bX)$. This is appropriate because $T$, when treated as a survival outcome of interest, is only subject to random censoring by $C$, and $I(\eta\neq 0)$ indicates whether $T$ is observed or not. 

When the interest pertains to  a specific type of failure, say type $k$,  we propose to consider the following variant of model \eqref{coxmodel} to define the causal treatment effect of interest:
\begin{equation} \label{coxmodel_k}
h_k(t; D, \xb) = h_{k, 0}(t) \exp\{\beta_{d, k} D + \bm{\beta}_{x, k}^{T} \xb\},
\end{equation}
where $h_k(t; D, \xb)$ is the type-$k$ cause-specific hazard function for compliers defined as
$$
h_k(t; D, \xb) = \lim_{\Delta t \rightarrow 0} \frac{\Pr(t \leq T \leq t + \Delta t| T \geq t, \delta=k, D_1 > D_0, D, \xb)}{\Delta t},
$$
and $h_{k, 0}(t)$ is an unspecified baseline cause-specific hazard at time $t$ for type $k$.  Under model \eqref{coxmodel_k}, $\bb_{d, k}$ represents the causal treatment effect on the type-$k$ cause-specific hazard for compliers after adjusting for covariates in $\bX$.  When all subjects are compliers, one can estimate model \eqref{coxmodel_k} using a slightly modified  partial likelihood score equation, which is \eqref{Ubasic1} with $I(\eta=k)$ replacing $\delta$ \citep{Kalbfleisch2011}. Following the same arguments for justifying the weighting technique presented in Section 2.3, we can show that incorporating $\hat\kappa$ or $\hat\kappa_v$ into this modified score equation yields an unbiased estimating equation for $\bb_{x, k}$. In other words, naively treating   the competing risks for type-$k$ failure as independent censoring events and applying the proposed IV method for randomly censored data lead to legitimate estimation and inference for the causal treatment effect on the type-$k$ cause-specific hazard.

{\bf Recurrent events}: In survival settings, the event of interest may occur repeatedly over time. The proportional hazards model can be naturally extended to a proportional intensity model to accommodate recurrent events \citep{Andersen1982}. Let $T^{(j)}$ denote the $j$-th recurrent event. Define $N^*(t)=\sum_{j=1}^\infty I(T^{(j)}\leq t)$ and $N^r(t)=\sum_{j=1}^\infty I(L<T^{(j)}\leq R)$, which respectively represent the underlying and the observed counting processes of recurrent events. Here $(L, R]$ denotes the time window in which recurrent events are observed. We assume $L$ and $R$ are independent of $V$ given $\bX$ and are independent of $T^{(j)}$'s conditional on $(V, D, \bX)$. Let $Y^r(t)=I(L<t\leq R)$, which denotes the at-risk process. A causal proportional intensity model is defined similarly to the Cox's proportional hazards model \eqref{coxmodel}:
\begin{equation}
\lambda(t)=\lambda_0(t) \exp\{\beta_{r, d} D + \bm{\beta}_{r, x}^{T} \xb\},
\end{equation}
where $\lambda(t)$ denotes the intensity function associated with $N^*(\cdot)$ given compliers (i.e. $D_1>D_0$), and $\lambda_0(t)$ is an unspecified baseline intensity function. The causal treatment effect on the recurrent events for compliers is captured by $\bb_{r,d}$.
As shown by \cite{Andersen1982},  in the setting where all subjects are compliers, $\bb_{r, d}$ can be estimated by equation \eqref{Ubasic1} with $N^r(\cdot)$ in place of $N(\cdot)$ and $Y^r(\cdot)$ in place of $Y(\cdot)$.  Adapting the weighting technique developed in Section 2.3 and 2.5, we can similarly modify the estimating equation for $\bb_{r, d}$ by incorporating weights $\hat\kappa$ or $\hat\kappa_v$. That is, we can obtain an unbiased estimate for $\bb_{r,d}$ by solving the  equation \eqref{proposed_ee} with $N^r(\cdot)$ and $Y^r(\cdot)$ in place of $N(\cdot)$ and $Y(\cdot)$.
}

\subsection{Large Sample Results}
\sloppy To ease presentation, we first introduce some new notation. Define
$$
\ubark(\bb)  = \frac{1}{\sqrt{n}}\sum_{i=1}^n  \int_0^{\infty}\kappa_i \left[ \bz_i  -  \left\{ \frac{ \bm{s}_{c}^{(1)}(\bb, s)} { s_{c}^{(0)}(\bb, s)}  \right\} \right] \,dM_i(s),
$$

$\bE_{n, \kappa}(\bb, t)=\frac{ \bS_{n, \kappa}^{(1)}(\bb, t)} {S_{n, \kappa}^{(0)}(\bb, t)}$, $\bE_{n, \hat\kappa}(\bb, t)=\frac{ \bS_{n, \hat\kappa}^{(1)}(\bb, t)} {S_{n, \hat\kappa}^{(0)}(\bb, t)}$, $\be_c(\bb, t)=\frac{ \bs_c^{(1)}(\bb, t)} {s_c^{(0)}(\bb, t)}$, $\bV_{n, \kappa}(\bb, t)=\frac{\bS_{n, \kappa}^{(2)}(\bb, t)} {S_{n, \kappa}^{(0)}(\bb, t)}-\bE_{n, \kappa}(\bb, t)^{\otimes 2}$, $\bV_{n, \hat\kappa}(\bb, t)=\frac{\bS_{n, \hat\kappa}^{(2)}(\bb, t)} {S_{n, \hat\kappa}^{(0)}(\bb, t)}-\bE_{n, \hat\kappa}(\bb, t)^{\otimes 2}$, and $\bv_c(\bb, t)=\frac{ \bs_c^{(2)}(\bb, t)} {s_c^{(0)}(\bb, t)}-\be_c(\bb, t)^{\otimes 2}$. 
Let $\bsig_0=\int_0^\infty \bv_c(\bb_0, t) s_c^{(0)}(\bb_0, t)h_0(t)dt$. We let $\|\cdot\|$ denote Euclidean norm.

We assume the following regularity conditions:

(C1): The parameter space for $\bb$, ${\boldsymbol{\cal B}}$, is compact.

(C2): $\|\bZ\|<\infty$ and $|\kappa|<\infty$.

(C3): $s_c^{(0)}(\bb, t)$ is bounded away from $0$ uniformly in $\bb$ and $t$.

(C4): $\bsig_0>0$.

(C5): $\hat\ab-\ab_0\rightarrow_{a.s.} 0$.

(C6) There exists an influence function $\bI_{\ab}(\cdot)$ such that $$
\|n^{1/2}(\ah - \an) - n^{-1/2} \sum_{i=1}^{n} I_{\ab}(\an, \bO_i)\|=o(1),\ a.s.
$$

We establish the consistency and the asymptotic normality for the proposed estimator in the following theorems:

\newtheorem{thm1}{Theorem}
\begin{thm1}\label{thm1}
	(Consistency) Under conditions (C1)-(C5),  $\bh \rightarrow_{a.s.} \bn$.
\end{thm1}

\begin{thm1}\label{thm2}
	(Asymptotic normality) Under conditions (C1)-(C6),  $n^{1/2}(\bh-\bn) \rightarrow_{d} N(0, \Omega)$, where $\Omega$ is defined in Appendix B (see equation \eqref{omega_matrix}).
\end{thm1}

The regularity conditions (C1)-(C2) impose the boundedness of the parameter space and covariates, which are mild and are often met in practice. The boundedness of $\kappa$ is satisfied when $\Pr(V=0|\bX)$ is always away from $0$ and $1$. Conditions (C3)-(C4) are standard assumptions for Cox proportional hazard regression methods. For example, condition (C4) ensures the identifiability of $\bb_0$. Conditions (C5)-C(6) depict reasonable requirements on the estimator of $\balpha_0$, such as consistency and i.i.d. sum representation.
The detailed proofs of Theorems \ref{thm1} and \ref{thm2} are provided in the Appendix B. %
%

\subsection{Variance Estimation}

In the proof of Theorem \ref{thm2}, we derive a closed form for the asymptotic variance of $n^{1/2}(\hat\bb-\bn)$; see equation \eqref{omega_matrix} in the Appendix B. A consistent variance estimator for $\hat\bb$ can be obtained by $\hat\bOmega/n$, where $\hat\bOmega$ is $\bOmega$ with unknown quantities replaced by their empirical counterparts or consistent estimators.

An alternative approach to estimating the asymptotic variance of $\hat\bb$ is to use bootstrapping. The detailed procedure follows:
\begin{enumerate}
	\item Resample $n$ observations from the original dataset with replacement, $\{\bO_i^{b}\}_{i=1}^{n}$, and add some small amount of noise (e.g. $N(0, 10^{-10})$) to avoid the presence of ties in the resampled data.
	\item  Calculate $\bh^b$ based on $\{\bO_i^b\}_{i=1}^n$ with weights as described in Section 2 (i.e. $\kh$, $\kh_v$, or $\kh_{v, tr}$).
	\item Repeat steps 1-2 for $b=1,...,B$.
	\item Estimate the asymptotic variance of $\bh$ by the empirical variance of $\{\bh^b\}_{b=1}^B$.
\end{enumerate}

In the bootstrapping procedure, the computations  in Step 2 may fail to converge. In such a case,  we would carry out Step 3 until there are $B$ convergent estimates. In addition, repeated resampling may occasionally produce outlier estimates that artificially inflate the empirical variance  in Step 4. When this occurs, one may estimate the standard deviation of $\hat\bb$ by the median absolute deviation, namely, $1.4826 \times MAD$, where $MAD = median( |\bh_b - median(\bh_b)|)$ \citep{Rousseeuw1993}.
This alternative approach performs quite well based on our numerical experiences.

\section{Simulation Study} \label{sec:simstudy}
We conduct extensive simulations to assess the performance of the proposed estimators. To create data under  assumptions (A1) to (A4), we take the following steps:
\begin{enumerate}
	\item Generate $\xb$ from some bounded distribution.
	\item Generate the latent group membership (i.e. \emph{complier}, \emph{always-taker} or \emph{never-taker}) from a multinomial distribution.
	\item Let $P(V=1|\xb) = \frac{\exp(\alpha_{01} + \alpha_{02}^T\xb)}{1 + \exp(\alpha_{01} + \alpha_{02}^T\xb)}$, and then generate $V \sim Bernoulli(P(V=1|\xb))$. Treatment $D$ is then automatically determined by $V$ and latent group membership generated in Step 2. 
	\item For compliers, generate potential survival times $T_{00} = T_{10} = \exp(-\bb_x^T\xb + \epsilon)$ and $T_{01}=T_{11} = \exp(-\bb_x^T\xb - \bb_d + \epsilon)$, where $\epsilon$ follows the extreme value distribution. 
	The survival time $T$ is then determined by $(V, D)$ (i.e. $T=T_{vd}$.  It is easy to show that $T$ follows the Cox regression model \eqref{coxmodel}.	
	\item For non-compliers, perform a similar process, where $T_{00} = T_{10}$ and $T_{01} = T_{11}$ and $T$ is determined by $(V, D)$ (i.e. $T=T_{vd}$). The model used to generate $T_{00}$ or $T_{01}$ is not restricted to the Cox regression model.
	\item Draw independent censoring times $C \sim Exponential(0.5)$.
\end{enumerate}

We consider two basic data generation scenarios with a single covariate $X$. In scenario 1, for compliers, survival times are generated  with $\bn = (\beta_d, \beta_x) = (-0.5, -0.2)$. 
Survival times for non-compliers in scenario 1 are generated according to $T = \exp(-0.02 X + \epsilon_1)$ where $\epsilon_1 \sim N(0, 0.01)$ (i.e. no treatment effect). In scenario 2, compliers' survival times are generated with $\bn=(-0.3, 0.05)$.
Non-compliers' survival times are also generated from a Cox proportional hazards regression model, where $T = \exp(0.5 D - 0.05 X + \epsilon_2)$ and $\epsilon_2$ follows the extreme value distribution.

For each scenario, $X$ is simulated from either a $Bernoulli(0.5)$ or $Uniform(-1,1)$ distribution, and we vary the probability of compliers  from $1/3$ to $2/3$ (with always-takers and never-takers sharing the remaining probability equally). We also vary the sample size between $n=1000$ and $n=4000$. We fix logistic regression parameters $(\alpha_{01}, \alpha_{02}) = (0, 1)$. Table 1 summarizes the 8 cases run for each of the 2 scenarios.
\begin{table}
	\caption{Cases considered for each scenario}
	
	\centering
	\begin{tabular}{cccc}
		\noalign{\smallskip}\hline\hline\noalign{\smallskip}
		Case & $P(D_1 > D_0)$ & $n$ & $X$ \\
		\hline
		1 & 1/3 & 1000 & Uniform(-1, 1) \\
		2 & 2/3 & 1000 & Uniform(-1, 1) \\
		3 & 1/3 & 4000 & Uniform(-1, 1) \\
		4 & 2/3 & 4000 & Uniform(-1, 1) \\
		5 & 1/3 & 1000 & Bernoulli(0.5) \\
		6 & 2/3 & 1000 & Bernoulli(0.5) \\
		7 & 1/3 & 4000 & Bernoulli(0.5) \\
		8 & 2/3 & 4000 & Bernoulli(0.5) \\
		\noalign{\smallskip}\hline\hline\noalign{\smallskip}
	\end{tabular}
\end{table}

We compare several different methods of estimation: (1) the \emph{benchmark} estimate based only on the compliers (unknown in a real data analysis); (2) the \emph{naive} estimate which assumes the entire sample follows the same Cox model; (3) the proposed $\kh$-weighted estimate; (4) the modified $\kh_v$-weighted estimate; (5) the estimate based on the truncated modified weights $\kh_{v,tr}$. Hereafter, we refer to these methods as ``$Complier$'', ``$\kappa$'', ``$\kappa_v$'', ``$\kappa_{v, tr}$''.

To estimate $\kh_v$ and $\kh_{v,tr}$, we estimate $v_0(U) = P(V=1|W, X, D, \delta)$ using the method described in Section 2.5 with a second-order logistic regression including the interaction between $W$ and $X$ for each of the 4 partitions by the censoring and treatment status. Without further mentioning, estimation using $\kh$ and $\kh_v$ follows the algorithms and caveats laid out in Section 2.4 and 2.5, where $\bh$ is estimated by maximizing the objective function in (\ref{objfun}). More specifically, we use the \texttt{R} function  \texttt{optim}  with the \texttt{BFGS}  method option \citep{rbase}, considering three different starting values (based on the naive estimate, $\pm 0.5$), to solve the maximization problem. {For the method $\kappa_{v, tr}$, we use the \texttt{R} function \texttt{coxph} to implement the proposed estimation as described in Section 2.5. For each method under comparison, we check whether the resulting estimate solves the proposed estimating equation 
	within some tolerance (e.g. $0.05$). We record a failure to converge if such an estimate cannot be produced. }

\begin{figure*}
	
	\includegraphics[width=0.98\textwidth]{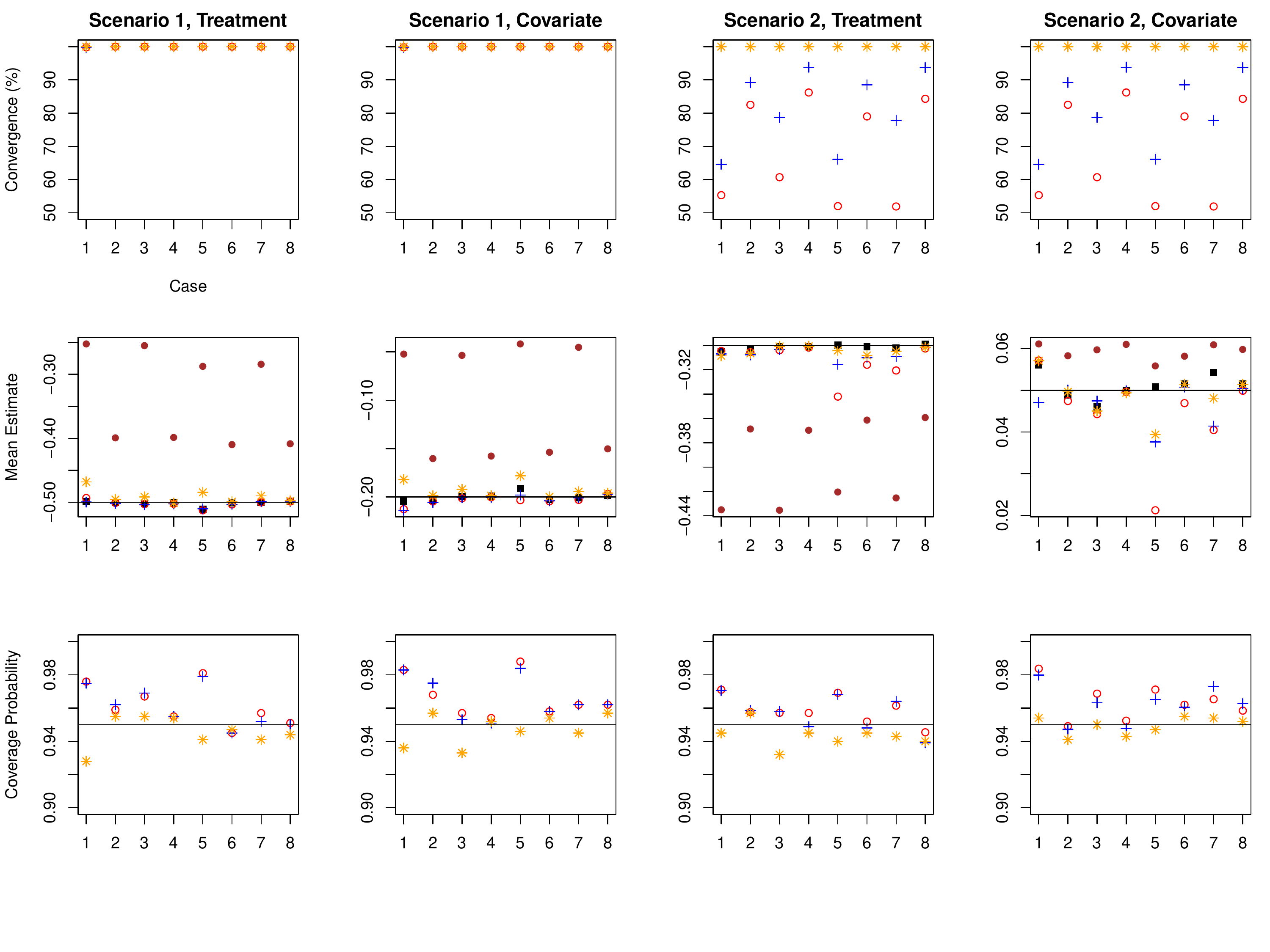}
	\caption{Simulation results: convergence rates, mean estimates, and empirical coverage probabilities of 95\% confidence intervals: Complier (\textcolor{black}{$\filledmedsquare$}); Naive (\textcolor{altbrown}{$\bullet$}); $\kappa$ (\textcolor{red}{$\circ$}); $\kappa_{v}$ (\textcolor{blue}{$+$}); $\kappa_{vtr}$ (\textcolor{orange}{\ding{83}})}
	\label{fig1}
\end{figure*}

The top row of Figure \ref{fig1} shows the convergence rates for the three proposed estimators. In scenario 1, the convergence rates of both $\kh$ and $\kh_v$ are close to  100\% times across the 8 cases considered. In scenario 2, the convergence rate  varies considerably, but generally increases with $n$ and the proportion of compliers $P(D_1 > D_0)$. Anecdotal examination reveals that the objective and estimating function surfaces for this scenario can be highly irregular. In contrast, as $\kh_{v,tr}$-weights are always positive, the resultant surfaces are smooth and the resulting convergence  rates are always 100\%.
The second row of Figure \ref{fig1} demonstrates the empirical bias by comparing  the treatment and covariate parameter estimates  to the truth. 
The naive parameter estimators generally demonstrate large empirical bias, with  the proposed methods reducing the bias considerably.

In Figure \ref{fig2}, we compare various standard error (SE) estimates to the empirical standard deviations (SD) of the proposed estimators. We denote the mean and median estimated SE based on the analytic variance estimation  by \emph{Mean SE} and \emph{Median SE} respectively, and denote the mean and median estimated SE based on the bootstrapping variance estimation by \emph{Mean Bootstrap SE} and \emph{Median Bootstrap SE} respectively. The empirical standard deviation (SD) is denoted by \emph{Empirical}.  
For the method $\kappa$, we evaluate both analytic variance estimation and bootstrapping based variance estimation.  It is observed that both {\emph{Mean Bootstrap SE} and \emph{Median Bootstrap SE} are rather close to the corresponding empirical SDs in both Scenarios 1 and 2.
	As for the analytic variance estimation,  \emph{Median SE}s are in good agreement with the empirical SDs,  while in Scenario 2, many \emph{Mean SE}s  considerably depart from the empirical SDs. The latter phenomenon may reflect the unstable performance of the $\kappa$-weighted estimator in Scenario 2, which is consistent with the lower convergence rates  of the method $\kappa$ in Scenario 2.
	For the methods $\kappa_v$ and $\kappa_{v,tr}$, we only examine the bootstrapping based variance estimation.
	Two extreme outliers are removed from calculating the mean bootstrap SE for the covariate coefficient estimate based on method $\kappa_v$ in the Case 5 of Scenario 1.
	We observe fairly small discrepancies among \emph{Mean Bootstrap SE}s, \emph{Median Bootstrap SE}s, and  empirical SDs for both Scenarios 1 and 2, while the method $\kappa_{v, tr}$ shows slightly better performance.
	
	The bottom row of Figure \ref{fig1} demonstrates the empirical coverage probabilities  of 95\% confidence intervals, constructed as $\bh \pm z_{0.975} \times \hat{SE}(\bh)$, where  $\hat{SE}(\bh)$ stands for the bootstrapping based SE.  The coverage probabilities associated with the method $\kappa_{v, tr}$ are fairly close to the nominal  95\% level,   dipping to 93\% in a few cases. The methods, $\kappa$ and $\kappa_v$, have similar and generally more conservative performance in terms of the empirical coverage probabilities. 
	Note that the results presented for these two methods are only based on simulations which produce converged estimates. In Scenario 2 where the convergence rates of $\kappa$ and $\kappa_v$ can be considerably below 1,  the results in Figure \ref{fig1} may over-represent the performance of these two methods.
	
	Based on all the simulations, the method $\kappa_{v,tr}$ evidences the best performance of the different weighting methods, exhibiting good coverage probabilities, low bias, and reliable convergence.

	\begin{figure*}
		
		\includegraphics[width=0.98\textwidth]{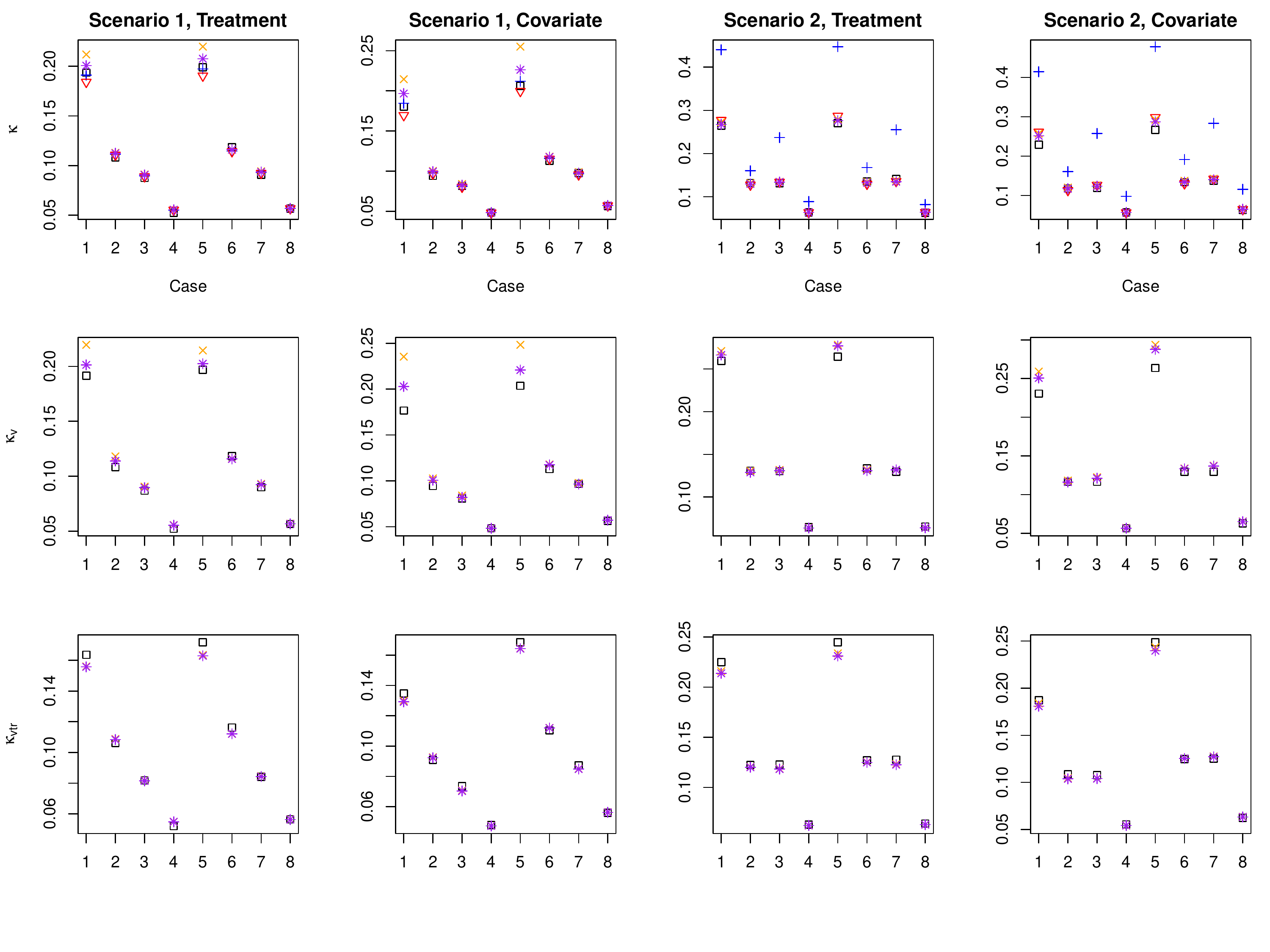}
		\caption{Simulation results: the estimated standard errors and empirical standard deviations of $\kh$, $\kh_v$, $\kh_{v, tr}$ weighted estimators:  Empirical (\textcolor{black}{$\medsquare$}); Mean SE (\textcolor{blue}{$+$}); Median SE (\textcolor{red}{$\medtriangledown$}); Mean Bootstrap SE (\textcolor{orange}{$\times$}); Median Bootstrap SE (\textcolor{violet}{\ding{83}})}
		\label{fig2}
	\end{figure*}

	%
	\clearpage
	
	\section{Colon Cancer Screening with Flexible Sigmoidoscopy}
	Colorectal cancer initiates in the colon or rectum (parts of the large intestine). It is the third most common cause of cancer deaths for both men and women in the United States \citep{acs}.
	Screening has been suggested for early detection of colon cancer and precancerous lesions known as polyps, with the ultimate goal of reducing colon cancer deaths. There are several recommended screening protocols, including fecal occult blood test (FOBT), fecal immunochemical test (FIT), colonoscopy, virtual colonoscopy and flexible  sigmoidoscopy.
	
	The Prostate, Lung, Colorectal, and Ovarian (PLCO) Cancer Screening Trial is a multi-center, two-armed  randomized trial, sponsored by the National Cancer Institute, of screening tests for prostate, lung, colorectal and ovarian cancers. Ten centers across the U.S. recruited approximately 155,000 participants between November 1993 and July 2001. Data were collected until December 31, 2009. One objective of the trial is evaluating the effectiveness of screening with flexible sigmoidoscopy on mortality from colorectal cancer compared to usual-care. \cite{prorok} reported further details about this trial.
	
	The original data consist of $154,897$ individuals aged $55$ to $74$ years. They were randomly assigned to either the usual-care (control, $N=77,453$) group or the screening with flexible sigmoidoscopy (intervention, $N=77,444$) group. For the intervention group, subjects were offered the screening at baseline and $3$ or $5$ years later.
	The data from 187 participants who dropped out, died, were diagnosed with cancer, or had an organ removed before the first screening visit and the data on 4 participants who have no follow-up after randomization are discarded.
	Thus, we only consider $154,706$ individuals in our analyses.
	
	Table \ref{tab:plco} presents descriptive statistics for the baseline characteristics of the participants stratified by the screening assignment (i.e. $V=0, V=1$) and the actual screening status (i.e. $D=0$, $D=1$). 
	We also consider risk factors, including age (in years), gender, family history of any cancer, family history of colorectal cancer, colorectal polyps, and diabetes.  We apply  t-tests or chi-square tests to check the balance of these observed risk factors between the groups determined by the screen assignment or the actual screening status.  Based on the p-values reported in Table \ref{tab:plco}, there is strong evidence that this trial was well randomized, with small and nonsignificant associations between the screening assignment and the risk factors. However, most of these risk factors are unbalanced by the actual screening status. The summary statistics in Table \ref{tab:plco} suggest that participants, who were older, male, with family history of any cancer or with family history of colorectal cancer,  or with diabetes, were more likely to take the colon cancer screening when it was assigned.  Thus, there is some evidence to suggest that  the study participants' post-randomization care selections and their potential survival outcomes are  dependent. Hence, the traditional ITT or the ``as-treated'' analysis may be problematic for evaluating the causal effect of flexible sigmoidoscopy screening on colorectal cancer mortality.
	
	\begin{table*}
		\caption{Characteristics of the Study Participants}
		\label{tab:plco}
		\centering
		\resizebox{\columnwidth}{!}{	
			\begin{tabular}{@{\extracolsep{5pt}}ll cc cc cc}
				\\[-1.8ex]\hline \hline \\[-1.8ex]
				\multicolumn{2}{l}{Characteristics} & Control ($V=0$) & Intervention ($V=1$) & & Not Screened ($D=0$) & Screened ($D=1$) & \\
				\noalign{\smallskip}
				&& $N=77449$ & $N=77257$ && $N=90056$ & $N=64650$&\\
				\noalign{\smallskip}\cmidrule(r){3-5}\cmidrule(r){6-8}\noalign{\smallskip}
				&& \multicolumn{2}{c}{\emph{Number of Participants} (\%)}& p-value &\multicolumn{2}{c}{\emph{Number of Participants} (\%)}& p-value\\
				\noalign{\smallskip}\hline\noalign{\smallskip}
				\multicolumn{2}{l}{Age $\star$} &&&&&& \\
				& & 62.60 (5.37) & 62.59 (5.39) & 0.8274& 62.65 (5.39) & 62.52 (5.33) &$<$.0001\\
				\noalign{\smallskip}
				\multicolumn{2}{l}{Age Level} &&&&&&\\
				& 55-59 yr & 25838 (33.36) & 25789 (33.38) && 29902 (33.20) & 21725 (33.60)&\\
				& 60-64 yr & 23767 (30.69) & 23736 (30.72) && 27451 (30.48) & 20052 (31.02)&\\
				& 65-69 yr & 17473 (22.56) & 17402 (22.52) && 20352 (22.60) & 14523 (22.46)&\\
				& 70-74 yr & 10371 (13.39) & 10330 (13.37) &0.9967& 12351 (13.71) & 8350 (12.92)&$<$.0001\\
				\noalign{\smallskip}
				\multicolumn{2}{l}{Sex} &&& &&&\\
				& Male & 38340 (49.50) & 38229 (49.48) && 43529 (48.34) & 33040 (51.11)&\\
				& Female & 39109 (50.50) & 39028 (50.52) &0.9393& 46527 (51.66) & 31610 (48.89)&$<$.0001\\
				\noalign{\smallskip}
				\multicolumn{2}{l}{Family History of Any Cancer} &&&&&& \\
				& No & 32742 (42.28) & 33327 (43.14) && 37798 (41.97) & 28271 (43.73)&\\
				& Yes & 41305 (53.33) & 41971 (54.33) &0.8735$\S$& 47137 (52.34) & 36139 (55.90)&0.0190$\S$\\
				& Unknown & 3402 (4.39) & 1959 (2.54) &$<$.0001& 5121 (5.69) & 240 (0.37)&$<$.0001\\
				\noalign{\smallskip}
				\multicolumn{2}{l}{Family History of Colorectal Cancer} &&& &&& \\
				& No & 64504 (83.29) & 65203 (84.40) && 73997 (82.17) & 55710 (86.17) &\\
				& Yes $\dag$ & 7320 (9.45) & 7627 (9.87) &0.0809$\S$& 8331 (9.25) & 6616 (10.23) &0.0022$\S$\\
				& Possibly $\ddag$/Unkown & 5625 (7.26) & 4427 (5.73) &$<$.0001& 7728 (8.58) & 2324 (3.59) &$<$.0001\\
				\noalign{\smallskip}
				\multicolumn{2}{l}{Colorectal Polyps} &&& &&& \\
				& No & 68690 (88.69) & 69910 (90.49) && 78705 (87.40) & 59895 (92.65)&\\
				& Yes & 4947 (6.39) & 5185 (6.71) &0.1565$\S$& 5739 (6.37) & 4393 (6.80)& 0.7865$\S$\\
				& Unknown & 3812 (4.92) & 2162 (2.80) &$<$.0001& 5612 (6.23) & 362 (0.56)&$<$.0001\\
				\noalign{\smallskip}
				\multicolumn{2}{l}{Diabetes} &&&&&&\\
				& No & 68028 (87.84) & 69371 (89.79) && 77773 (86.36) & 59626 (92.23)&\\
				& Yes & 5699 (7.36) & 5810 (7.52) &0.9971$\S$& 6776 (7.52) & 4733 (7.32)&$<$.0001$\S$\\
				& Unknown & 3722 (4.81) & 2076 (2.69) &$<$.0001& 5507 (6.12) & 291 (0.45)&$<$.0001\\
				\noalign{\smallskip}\hline\noalign{\smallskip}
				\multicolumn{6}{l}{$\star$ denotes a continuous variable. Mean and standard deviation are reported.}\\
				\multicolumn{6}{l}{$\dag$ indicates colorectal cancer family history in immediate family member.}\\
				\multicolumn{6}{l}{$\ddag$ indicates colorectal cancer family history in relatives or unclear cancer type.}\\
				\multicolumn{6}{l}{$\S$ indicates p-value without considering missing category.}\\
				\noalign{\smallskip}\hline\hline\noalign{\smallskip}
			\end{tabular}
		}
	\end{table*}
	
	To address this issue, we employ the proposed IV methods, with the survival outcome of interest ($T$) defined as the time from trial entry (i.e. randomization) to death from colorectal cancer (in years), and
	the IV chosen as the screening assignment ($V$).  {In our dataset, $351$ and $249$ colorectal cancer deaths were observed in the control group ($n=77098$) and the intervention group ($n=77,098$)  respectively;  409 and 191 colorectal cancer deaths were observed in the group without screening ($n=89,647$) and the group with screening ($n=64,459$) respectively. In our analysis, deaths due to other causes are competing risks for death from colon cancer. As discussed in Section 2.6, naively treating such competing events as censoring events leads to a valid IV proportional hazards analysis of the cause-specific hazard function for colon cancer death. Our instrumental variable is justified as follows: (i) the screening assignment is highly informative of the actual screening status ($D$) (i.e. screened vs. not screened); (ii) the screen assignment is random and hence is expected to be independent of unmeasured confounders (given the observed risk factors); (iii) it is reasonable to expect that the  impact of the screening assignment on the survival outcome is only through its influence on the actual screening status.
	
	It is worth noting that this study assumes that individuals who were assigned usual care did not have access to other colorectal screening programs (i.e. $P(D_{0}=0|\bm{X})=1$). This renders a special case of perfect exclusion of the control and treatment groups \citep{Abadie2003}, where
	the assumption (A4) holds trivially. 
	In this case, we have
	\[\lambda(T | D_{1}>D_{0}, \bm{X}, D=1)=\lambda(T_{1} | D_{1}=1, \bm{X}, V=1)=\lambda(T_{1} | D=1, \bm{X})\]
	and
	\begin{align*}
		\begin{split}
			\lambda(T | D_{1}>D_{0}, \bm{X}, D=0)=\lambda(T_{0} | D_{1}=1, \bm{X}, V=0)\\
			=\lambda(T_{0} | D_{1}=1, \bm{X}, V=1)=\lambda(T_{0} | D=1, \bm{X}).
		\end{split}
	\end{align*}
	Thus, $\beta_d$ in our IV analyses  can be interpreted as the causal effect of the flexible sigmoidoscopy screening for screened participants given observed risk factors.

	\begin{table*}
		\caption{Analyses for the Unadjusted Screening Effect Based on the Whole Data Set or Stratified by Each Risk Factor}
		\label{tab:anal1}
		\centering
		\scriptsize
		\resizebox{\columnwidth}{!}{
			\begin{tabular}{@{\extracolsep{5pt}}l l l ccccc}
				\\[-1.8ex]\hline \hline \\[-1.8ex] \noalign{\smallskip}
				Data & $N$ & ${p}_{c}$ & As-Treated & ITT & $\kappa$ & $\kappa_{v}$ & $\kappa_{v,tr}$\\
				\noalign{\smallskip}\cmidrule(r){4-8}
				(Subgroup) & & & \multicolumn{5}{c}{Parameter Estimates (Standard Errors)}\\
				\noalign{\smallskip}\hline\noalign{\smallskip}
				Total & 154706 & 0.84 & -0.442* & -0.343* & -0.427* & -0.427* & -0.427*\\
				&&& (0.088) & (0.083) & (0.099) & (0.097) & (0.101)\\
				\noalign{\smallskip}
				Age Level \\
				55-59 yr & 51627 & 0.84 & -0.572* & -0.380* & -0.496* & -0.496* & -0.496*\\
				&&& (0.198) & (0.184) & (0.229) & (0.240) & (0.248)\\
				60-64 yr & 47503 & 0.84 & -0.313 & -0.130 & -0.169 & -0.169 & -0.169\\
				&&& (0.160) & (0.153) & (0.201) & (0.198) & (0.193)\\
				65-69 yr & 34875 & 0.83 & -0.475* & -0.590* & -0.654* & -0.655* & -0.655*\\
				&&& (0.164) & (0.158) & (0.178) & (0.164) & (0.182)\\
				70-74 yr & 20701 & 0.81 & -0.420* & -0.264 & -0.351 & -0.350 & -0.350\\
				&&& (0.188) & (0.176) & (0.228) & (0.213) & (0.218)\\
				\noalign{\smallskip}
				Sex\\
				Male & 76569 & 0.86 & -0.549* & -0.445* & -0.536* & -0.536* & -0.536*\\
				&&& (0.115) & (0.109) & (0.124) & (0.123) & (0.123)\\
				Female & 78137 & 0.81 & -0.319* & -0.200 & -0.262 & -0.262& -0.262\\
				&&& (0.135) & (0.128) & (0.156) & (0.172) & (0.166)\\
				\noalign{\smallskip}
				Family History of Any Cancer \\
				Yes & 83276 & 0.86 & -0.237* & -0.258* & -0.294* & -0.294* & -0.294* \\
				&&& (0.114) & (0.111) & (0.120) & (0.124) & (0.127)\\
				No & 66069 & 0.85 & -0.704* & -0.492* & -0.639* & -0.639* & -0.639*\\
				&&& (0.144) & (0.132) & (0.158) & (0.179) & (0.162)\\
				\noalign{\smallskip}
				Family History of Colorectal Cancer\\
				Yes & 14947 & 0.87 & -0.010 & -0.097 & -0.105 & -0.106 & -0.106\\
				&&& (0.241) & (0.239) & (0.251) & (0.271) & (0.254) \\
				No & 129707 &  0.85 & -0.457* & -0.391* & -0.469* & -0.469* & -0.469*\\
				&&& (0.099) & (0.094) & (0.117) & (0.113) & (0.104)\\
				\noalign{\smallskip}
				Colorectal Polyps\\
				Yes & 10132 & 0.85 & 0.315 & 0.288 & 0.335 & 0.336 & 0.336\\
				&&& (0.305) & (0.309) & (0.388) & (0.405) & (0.389)\\
				No & 138600 & 0.86 & -0.490* & -0.401* & -0.490* & -0.485* & -0.485*\\
				&&& (0.093) & (0.089) & (0.111) & (0.112) & (0.110)\\
				\noalign{\smallskip}
				Diabetes\\
				Yes & 11509 & 0.81 & -1.036* & -0.335 & -0.606 & -0.603 & -0.603\\
				&&& (0.311) & (0.253) & (0.451) & (0.454) & (0.438)\\
				No & 137399 & 0.86 & -0.355* & -0.349* & -0.404* & -0.404* & -0.404*\\
				&&& (0.093) & (0.090) & (0.095) & (0.099) & (0.092)\\
				\noalign{\smallskip}\hline\noalign{\smallskip}
				\multicolumn{6}{l}{* indicates $p$-value $\leq$ 0.05}\\
				\noalign{\smallskip}\hline\hline\noalign{\smallskip}
			\end{tabular}
		}
	\end{table*}
	
	We first assess the unadjusted causal effect of the flexible sigmoidoscopy screening by fitting model \eqref{coxmodel} without $\bX$ to the full dataset and stratifying the analysis by each risk factor.
	For comparison purposes, we also perform the ``as-treated'' counterparts (i.e. fitting a Cox model for $T$ with $D$ being the only covariate), and the ITT counterparts (i.e. fitting a Cox model for $T$ with $V$ being the only covariate) of these IV analyses.
	For the IV analyses, we implement the three methods $\kappa$, $\kappa_v$, and $\kappa_{v, tr}$ in the same way as in our simulation studies (see Section \ref{sec:simstudy}), {except we use a simple logistic regression model stratified by $(\delta, D)$ to estimate the $v_0(U)$ in \eqref{modkappa}.}
	Table \ref{tab:anal1} reports the parameter estimates and the associated standard errors. {For the IV methods, we present the bootstrap-based standard errors.}

	Table \ref{tab:anal1} also reports the rate of compliance in the intervention group (i.e. the proportion of screened participants in the intervention group), $p_{c}$. 
	
	From Table \ref{tab:anal1}, we observe that the estimates for the causal effect of screening are very similar among the three IV methods. The conclusions regarding the survival impact of screening are generally consistent across the IV analyses, the as-treated analyses, and the ITT analyses, except for the sub-cohort with baseline age between 70 and 74 years and the sub-cohort with diabetes. In these two cases, rather large, significant benefits of screening are suggested by the as-treated analyses but not by  the ITT or IV analyses. Such discrepancies may be explained by the relatively high noncompliance rates ($\approx 19\%$) observed in the intervention group. That is, study participants who refused assigned screening are likely to be less health-conscious, which may be associated with worse potential survival outcomes.  When the non-screened group includes a large proportion of such participants, the as-treated analyses would tend to over-estimate the benefit of screening as a result of ignoring the survival  impact of the unmeasured confounder related to health-consciousness. Therefore, in these two cases, it is more plausible  to conclude that  the flexible sigmoidoscopy screening offers little survival benefits for the participants aged between 70 and 74 years and for participants with diabetes. Overall, the unadjusted stratified analyses support the benefit of flexible sigmoidoscopy in reducing colorectal mortality, with the greatest benefit in subpopulations with relatively low mortality risk, for example, age group 55-59 years and subjects without family history of colorectal cancer.
	
\begin{sidewaystable*}
		\caption{Characteristics of the Study Participants by Age Subgroups}
		\label{tab:plco_age}
		\centering
		\resizebox{\columnwidth}{!}{
			\begin{tabular}{@{\extracolsep{5pt}}l cccccc cccccc}
				\\[-1.8ex]\hline \hline \\[-1.8ex]
				Age Level& \multicolumn{6}{c}{55-59 yr} & \multicolumn{6}{c}{60-64 yr}\\
				\noalign{\smallskip}\cmidrule(r){2-7}\cmidrule(r){8-13}
				Covariates &$V=0$ & $V=1$ & $p$-value & $D=0$ & $D=1$ & $p$-value &$V=0$ & $V=1$ & $p$-value & $D=0$ & $D=1$ & $p$-value\\
				\noalign{\smallskip}\hline\noalign{\smallskip}
				Gender&\\
				\hspace{1pc}Male & 11078 (46.8)	&	11576 (47.6)	&&	12403 (46.1)	&	10251 (48.7)	&&10831 (49.5)	&	11145 (50.0)	&&	12074 (48.4)	&	9902 (51.4)	&\\
				\hspace{1pc}Female & 12595 (53.2)	&	12724 (52.4)	&0.0661&	14530 (53.9)	&	10789 (51.3)	&$<$.0001*&11070 (50.5)	&	11164 (50.0)	&0.2946&	12886 (51.6)	&	9348 (48.6)	&$<$.0001*\\
				Family History of Any Cancer &\\
				\hspace{1pc}No & 11148 (47.1)	&	11545 (47.5)	&&	12793 (47.5)	&	9900 (47.1)	&&9979 (45.6)	&	10139 (45.4)	 &&	11427 (45.8)	&	8691 (45.1)	&\\
				\hspace{1pc}Yes & 12525 (52.9)	&	12755 (52.5)	&0.3633 &	14140 (52.5)	&	11140 (52.9)	&0.3361&11922 (54.4)	 &	12170 (54.6)	&0.8138&	13533 (54.2)	&	10559 (54.9)	&0.1882\\
				Family History of Colorectal Cancer&\\
				\hspace{1pc}No & 21485 (90.8)	&	22000 (90.5)	&&	24455 (90.8)	&	19030 (90.4)	&&19681 (89.9)	&	19923 (89.3)	&&	22455 (90.0)	&	17149 (89.1)	&\\
				\hspace{1pc}Yes & 2188 (9.2)	&	2300 (9.5)	&0.4118&	2478 (9.2)	&	2010 (9.6)	&0.1935&2220 (10.1)	&	2386 (10.7)	 &0.0565&	2505 (10.0)	&	2101 (10.9)	&0.0029*\\
				Colorectal Polyps &\\
				\hspace{1pc}No & 22691 (95.9)	&	23264 (95.7)	&&	25806 (95.8)	&	20149 (95.8)	&&20432 (93.3)	&	20747 (93.0)	&&	23259 (93.2)	&	17920 (93.1)	&\\
				\hspace{1pc}Yes & 982 (4.1)	&	1036 (4.3)	&0.5448&	1127 (4.2)	&	891 (4.2)	&0.8029&1469 (6.7)	&	1562 (7.0)	 &0.2282&	1701 (6.8)	&	1330 (6.9)	&0.7117\\
				Diabetes &\\
				\hspace{1pc}No & 22217 (93.8)	&	22888 (94.2)	&&	25223 (93.7)	&	19882 (94.5)	&&20243 (92.4)	&	20648 (92.6)	&&	23008 (92.2)	&	17883 (92.9)	&\\
				\hspace{1pc}Yes & 1456 (6.2)	&	1412 (5.8)	&0.1211&	1710 (6.3)	&	1158 (5.5)	&0.0001*&1658 (7.6)	&	1661 (7.4)	 &0.6308&	1952 (7.8)	&	1367 (7.1)	&0.0047*\\
				\noalign{\smallskip}\hline\noalign{\smallskip}
				&\multicolumn{6}{c}{65-69 yr} & \multicolumn{6}{c}{70-74 yr}\\
				\noalign{\smallskip}\cmidrule(r){2-7}\cmidrule(r){8-13}
				&$V=0$ & $V=1$ & $p$-value & $D=0$ & $D=1$ & $p$-value &$V=0$ & $V=1$ & $p$-value & $D=0$ & $D=1$ & $p$-value\\
				\noalign{\smallskip}\hline\noalign{\smallskip}
				Gender&\\
				\hspace{1pc}Male & 8042 (50.1)	&	8192 (50.4)	&&	8985 (48.7)	&	7249 (52.3)	&& 4579 (48.1)	&	4641 (48.5)	&&	5177 (46.3)	&	4043 (51.0)\\
				\hspace{1pc}Female & 8015 (49.9)	&	8073 (49.6)	&0.6203&	9483 (51.3)	&	6605 (47.7)	&$<$.0001*&4949 (51.9)	&	 4925 (51.5)	&0.5368&	5997 (53.7)	&	3877 (49.0) & $<$.0001*\\
				Family History of Any Cancer &\\
				\hspace{1pc}No & 7252 (45.2)	&	7309 (44.9)	&&	8399 (45.5)	&	6162 (44.5)	&&4083 (42.9)	&	4153 (43.4)	&&	4844 (43.4)	&	3392 (42.8)&\\
				\hspace{1pc}Yes & 8805 (54.8)	&	8956 (55.1)	&0.6898&	10069 (54.5)	&	7692 (55.5)	&0.0754&4083 (57.1)	&	4153 (56.6)	&0.4421&	4844 (56.6)	&	3392 (57.2)&0.4819\\
				Family History of Colorectal Cancer&\\
				\hspace{1pc}No & 14320 (89.2)	&	14461 (88.9)	&&	16476 (89.2)	&	12305 (88.8)	&&4083 (88.4)	&	4153 (88.5)	 &&	4844 (88.6)	&	3392 (88.2)\\
				\hspace{1pc}Yes & 1737 (10.8)	&	1804 (11.1)	&0.4415&	1992 (10.8)	&	1549 (11.2)	&0.2686&1104 (11.6)	&	1102 (11.5)	 &0.9029&	1275 (11.4)	&	931 (11.8)&0.4771\\
				Colorectal Polyps &\\
				\hspace{1pc}No & 1737 (91.5)	&	1804 (91.2)	&&	1992 (91.5)	&	1549 (91.3)	&&8561 (89.9)	&	8582 (89.7)	&&	10036 (89.8)	&	7107 (89.7)\\
				\hspace{1pc}Yes & 1360 (8.5)	&	1426 (8.8)	&0.3509&	1576 (8.5)	&	1210 (8.7)	&0.5387&967 (10.1)	&	984 (10.3)	 &0.7722&	1138 (10.2)	&	813 (10.3)&0.8750\\
				Diabetes &\\
				\hspace{1pc}No & 14652 (91.2)	&	14799 (91.0)	&&	16798 (91.0)	&	12653 (91.3)	&&967 (90.5)	&	984 (89.5)	 &&	1138 (90.1)	&	813 (89.8)\\
				\hspace{1pc}Yes & 1405 (8.8)	&	1466 (9.0)	&0.4169&	1670 (9.0)	&	1201 (8.7)	&0.2506&907 (9.5)	&	1007 (10.5)	 &0.0218*&	1110 (9.9)	&	804 (10.2) & 0.6390\\
				\noalign{\smallskip}\hline\noalign{\smallskip}
				\multicolumn{6}{l}{$N$ (Row Percentage, \%)}\\
				\multicolumn{6}{l}{* indicates $p$-value $\leq$ 0.05}\\
				\noalign{\smallskip}\hline\hline\noalign{\smallskip}
			\end{tabular}
		}
\end{sidewaystable*}	
		
	\begin{table*}
		\caption{Results of Adjusted Models within Age Subgroups}
		\label{tab:anal2}
		\centering
		\resizebox{\columnwidth}{!}{
			\begin{tabular}{@{\extracolsep{5pt}}ll ccccc}
				\\[-1.8ex]\hline \hline \\[-1.8ex]
				Age Level && As-Treated & ITT & $\kappa$ & $\kappa_{v}$ & $\kappa_{v,tr}$\\
				\noalign{\smallskip}\cmidrule(r){3-7}
				(${p}_{c}$)&Covariates & \multicolumn{5}{c}{Point Estimates (Standard Errors)}\\
				\noalign{\smallskip}\hline\noalign{\smallskip}
				55-59 yr & Screening & -0.474* (0.207) & -0.296 (0.196) & -0.373 (0.228) & -0.373 (0.242) & -0.373 (0.246)\\
				(0.84) & Female & -0.101 (0.195) & -0.089 (0.195) & -0.003 (0.244) & -0.013 (0.246) & -0.013 (0.232)\\
				& Family History of Any Cancer & 0.204 (0.208) & 0.201 (0.208) & 0.468 (0.272) & 0.463 (0.280) & 0.465 (0.280)\\
				& Family History of Colorectal Cancer & 0.194 (0.313) & 0.192 (0.313) & -0.080 (0.471) & -0.071 (0.468) & -0.073 (0.386)\\
				& Colorectal Polyps & 0.276 (0.422) & 0.277 (0.422) & 0.137 (0.736) & 0.135 (1.725) & 0.131 (1.768)\\
				& Diabetes & 0.168 (0.392) & 0.179 (0.392) & 0.127 (0.606) & 0.125 (0.591) & 0.126 (0.710)\\
				\noalign{\smallskip}\hline\noalign{\smallskip}
				60-64 yr & Screening & -0.333* (0.169) & -0.184 (0.163) & -0.228 (0.197) & -0.229 (0.205) & -0.242 (0.181)\\
				(0.84)& Female & -0.419* (0.167) & -0.409* (0.166) & -0.579* (0.214) & -0.585* (0.206) & -0.563* (0.189)\\
				& Family History of Any Cancer & -0.182 (0.176) & -0.183 (0.176) & -0.055 (0.234) & -0.054 (0.231) & -0.071 (0.225)\\
				& Family History of Colorectal Cancer & 0.396 (0.260) & 0.391 (0.260) & 0.564* (0.279) & 0.566* (0.275) & 0.556* (0.276)\\
				& Colorectal Polyps & -0.124 (0.329) & -0.121 (0.329) & -0.141 (0.429) & -0.147 (0.446) & -0.108 (0.351)\\
				& Diabetes & 0.520* (0.258) & 0.526* (0.258) & 0.114 (0.458) & 0.117 (0.554) & 0.206 (0.369)\\
				\noalign{\smallskip}\hline\noalign{\smallskip}
				65-69 yr & Screening & -0.386* (0.168) & -0.526* (0.165) & -0.564* (0.187) & -0.568* (0.166) & -0.576* (0.188)\\
				(0.83)& Female & -0.402* (0.164) & -0.388* (0.164) & -0.426* (0.194) & -0.435* (0.181) & -0.408* (0.182)\\
				& Family History of Any Cancer & -0.182 (0.176) & -0.185 (0.176) & -0.187 (0.190) & -0.190 (0.196) & -0.198 (0.186)\\
				& Family History of Colorectal Cancer & 0.565* (0.129) & 0.563* (0.139) & 0.625* (0.260) & 0.642* (0.242) & 0.627* (0.253)\\
				& Colorectal Polyps & -0.306 (0.314) & -0.299 (0.314) & -0.226 (0.331) & -0.243 (0.349) & -0.245 (0.328)\\
				& Diabetes & 0.370 (0.251) & 0.377 (0.251) & 0.036 (0.419) & 0.045 (0.411) & 0.138 (0.313)\\
				\noalign{\smallskip}\hline\noalign{\smallskip}
				70-74 yr & Screening & -0.414* (0.196) & -0.364 (0.186) & -0.437 (0.225) & -0.439* (0.223) & -0.439* (0.223)\\
				(0.81)& Female & -0.486* (0.189) & -0.467* (0.189) & -0.472* (0.228) & -0.484* (0.246) & -0.486* (0.228)\\
				& Family History of Any Cancer & 0.157 (0.195) & 0.152 (0.195) & 0.387 (0.250) & 0.389 (0.268) & 0.388 (0.233)\\
				& Family History of Colorectal Cancer & -0.219 (0.318) & -0.222 (0.318) & -0.344 (0.508) & -0.333 (0.405) & -0.342 (0.384)\\
				& Colorectal Polyps & 0.088 (0.286) & 0.093 (0.286) & 0.137 (0.387) & 0.124 (0.396) & 0.121 (0.349)\\
				& Diabetes & 0.444 (0.270) & 0.451 (0.270) & -0.027 (0.583) & -0.031 (0.589) & -0.037 (0.409)\\
				\noalign{\smallskip}\hline\noalign{\smallskip}
				\multicolumn{6}{l}{* indicates $p$-value $\leq$ 0.05}\\
				\noalign{\smallskip}\hline\hline\noalign{\smallskip}
			\end{tabular}
		}
	\end{table*}
	
	We next evaluate the causal effect of screening while accounting for other risk factors.  Specifically, we fit model \eqref{coxmodel} with $\bX$ capturing gender, family history of any cancer, family history of colorectal cancer, colorectal polyps, and diabetes,  separately for the four age groups, 55-59 years, 60-64 years, 65-69 years, and 70-74 years.
	Table \ref{tab:plco_age} provides the summary statistics (i.e. count and percentage) of the risk factors by $V$ and by $D$ within each age group, along with the p values from testing the association of the risk factors with $V$ or $D$ based on the Chi-square tests. Similarly to Table \ref{tab:plco}, within each age group, the risk factors show little association with the screening assignment $D$ but may be significantly different between the participants who were screened versus those who were not screened.
	
	Table \ref{tab:anal2} presents the parameter estimates and  the associated standard errors based on the IV methods, $\kappa$, $\kappa_v$, and $\kappa_{v, tr}$. The coefficient estimates from the as-treated analysis (i.e. a multivariate Cox model for $T$ given $D$ and $\bX$) and the ITT analysis (i.e. a multivariate Cox model for $T$ given $V$ and $\bX$) are also presented along with the corresponding standard errors. From Table \ref{tab:anal2}, we again observe a quite good agreement among the three IV estimates. The IV analyses suggest that the flexible sigmoidoscopy screening has a significant protective effect on colorectal cancer mortality in the older age groups, such as 65-69 years and 70-74 years, but not in the younger age groups, 55-59 years and 60-64 years, after adjusting for age, gender, family history of any cancer, family history of colorectal cancer, and colorectal polyps, and diabetes. This finding is generally consistent with that based on the ITT analyses, but moderately disagrees with the results from the as-treated analyses, particularly in the age groups, 55-59 years and 60-64 years. In these two age groups,  we observe a more marked imbalance of risk factors by the actual screening status, compared to that presented in the two older age groups. For example, in the age group 60-64 years,
	participants who were female, had diabetes, or had no family history of colorectal cancer are significantly less likely to comply to the assigned screening assignment than those who were males, had no diabetes, or had a family of colorectal cancer.  Such associations may bias the estimation of the causal treatment effect by the as-treated analyses, and this may explain the discrepancies observed in Table \ref{tab:anal2} between the as-treated analyses and the IV analyses. In addition, the IV analyses provide strong evidence for the lower colorectal cancer mortality risk in females (versus males)  in all age groups beyond the age of 60 years. They also suggest some survival disadvantages (regarding colorectal cancer mortality) associated with the presence of family history of colorectal cancer.	
	
	\section{Concluding Remarks}
	The use of instrumental variables in survival settings with binary treatments has been severely limited by complexities arising from nonlinear model specifications, as with the proportional hazards model. The application of simple two stage estimation procedures developed for linear models is challenging and only valid in special cases. Alternative procedures may entail strong modelling assumptions on strata other than that of interest, tend to be complex, both computationally and inferentially, and are not readily implemented using standard software. Our approach based on a special characterization of instrumental variables enables a simple two stage procedure analogous to propensity score weighting. At the first stage, a binary regression model is fit to the instrumental variable while at the second stage, the fitted regression model from the first stage is used to construct a weight which ``debiases'' naive estimating equation for the proportional hazards model. Previous work  on this approach \citep{Abadie2002, Abadie2003} has only considered iid estimating equations with limited attention to the practical computational issues. The current paper demonstrates rigorously its validity with the partial likelihood score function. Moreover, the proposed estimators can be easily computed using existing software for the proportional hazards model, with variance estimation based on bootstrapping correctly accounting for the first stage estimation of the weights.

	{As shown in Section 2.6, the weighting approach is generally applicable to instrumental variable estimation of proportional hazards model in complex survival scenarios, for example, in the presence of left truncation, competing risks, and recurrent events.} The developed weighting technique may be applied to other survival regression models with binary treatment and binary instrumental variable. Application to the quantile regression model, the accelerated lifetime model, the additive hazard model, and transformation models are currently under investigation with right censored survival data. The main requirement is the existence of an unbiased estimating equation in the absence of unmeasured confounding. Such estimating equations can be incorporated into the second stage of the two stage procedure described in this paper for the proportional hazards model.
	
	
	\section*{Acknowledgements}
	
	The authors would like to express special thanks to Jerome Mabie, Tom Riley, Ryan
	Nobel and Josh Rathmell, Information Management Services (IMS) Inc, for supporting and
	managing the PLCO data. The authors also thank Dr. Stuart G. Baker, National Cancer
	Institute, for kindly introducing the IMS team for this research. The authors gratefully acknowledge the support from the National Institutes of Health grant R01 HL113548.

	\bibliography{ivcoxreg}

\begin{thebibliography}{31}
\providecommand{\natexlab}[1]{#1}
\providecommand{\url}[1]{\texttt{#1}}
\expandafter\ifx\csname urlstyle\endcsname\relax
  \providecommand{\doi}[1]{doi: #1}\else
  \providecommand{\doi}{doi: \begingroup \urlstyle{rm}\Url}\fi

\bibitem[Abadie(2003)]{Abadie2003}
A.~Abadie.
\newblock Semiparametric instrumental variable estimation of treatment response
  models.
\newblock \emph{Journal of econometrics}, 113\penalty0 (2):\penalty0 231--263,
  2003.

\bibitem[Abadie et~al.(2002)Abadie, Angrist, and Imbens]{Abadie2002}
A.~Abadie, J.~Angrist, and G.~Imbens.
\newblock Instrumental variables estimates of the effect of subsidized training
  on the quantiles of trainee earnings.
\newblock \emph{Econometrica}, 70\penalty0 (1):\penalty0 91--117, 2002.

\bibitem[Agresti(2013)]{Agresti2013}
A.~Agresti.
\newblock \emph{Categorical Data Analysis}.
\newblock Wiley Series in Probability and Statistics. Wiley, 2013.
\newblock ISBN 9780470463635.
\newblock URL \url{https://books.google.com/books?id=UOrr47-2oisC}.

\bibitem[Andersen and Gill(1982)]{Andersen1982}
P.~K. Andersen and R.~D. Gill.
\newblock Cox's regression model for counting processes: A large sample study.
\newblock \emph{The Annals of Statistics}, 10\penalty0 (4):\penalty0
  1100--1120, 1982.

\bibitem[Andersen et~al.(2012)Andersen, Borgan, Gill, and
  Keiding]{Andersen2012}
P.~K. Andersen, O.~Borgan, R.~D. Gill, and N.~Keiding.
\newblock \emph{Statistical models based on counting processes}.
\newblock Springer Science \& Business Media, 2012.

\bibitem[Angrist and Imbens(1995)]{Angrist1995}
J.~D. Angrist and G.~W. Imbens.
\newblock Two-stage least squares estimation of average causal effects in
  models with variable treatment intensity.
\newblock \emph{Journal of the American Statistical Association}, 90:\penalty0
  431--442, 1995.

\bibitem[Angrist et~al.(1996)Angrist, Imbens, and Rubin]{Angrist1996}
J.~D. Angrist, G.~W. Imbens, and D.~B. Rubin.
\newblock Identification of causal effects using instrumental variables.
\newblock \emph{Journal of the American Statistical Association}, 91\penalty0
  (434):\penalty0 444--455, 1996.

\bibitem[Baiocchi et~al.(2014)Baiocchi, Cheng, and Small]{Baiocchi2014}
M.~Baiocchi, J.~Cheng, and D.~S. Small.
\newblock Instrumental variable methods for causal inference.
\newblock \emph{Statistics in medicine}, 33\penalty0 (13):\penalty0 2297--2340,
  2014.

\bibitem[Baker(1998)]{Baker1998}
S.~G. Baker.
\newblock Analysis of survival data from a randomized trial with all-or-none
  compliance: estimating the cost-effectiveness of a cancer screening program.
\newblock \emph{Journal of the American Statistical Association}, 93\penalty0
  (443):\penalty0 929--934, 1998.

\bibitem[Cuzick et~al.(2007)Cuzick, Sasieni, Myles, and Tyrer]{Cuzick2007}
J.~Cuzick, P.~Sasieni, J.~Myles, and J.~Tyrer.
\newblock Estimating the effect of treatment in a proportional hazards model in
  the presence of non-compliance and contamination.
\newblock \emph{Journal of the Royal Statistical Society: Series B (Statistical
  Methodology)}, 69\penalty0 (4):\penalty0 565--588, 2007.

\bibitem[Gourieroux and Monfort(1981)]{Gourieroux1981}
C.~Gourieroux and A.~Monfort.
\newblock Asymptotic properties of the maximum likelihood estimator in
  dichotomous logit models.
\newblock \emph{Journal of Econometrics}, 17\penalty0 (1):\penalty0 83--97,
  1981.

\bibitem[Holland(1986)]{Holland1986}
P.~W. Holland.
\newblock Statistics and causal inference.
\newblock \emph{Journal of the American statistical Association}, 81\penalty0
  (396):\penalty0 945--960, 1986.

\bibitem[Joffe(2001)]{Joffe2001}
M.~M. Joffe.
\newblock Administrative and artificial censoring in censored regression
  models.
\newblock \emph{Statistics in medicine}, 20\penalty0 (15):\penalty0 2287--2304,
  2001.

\bibitem[Kalbfleisch and Prentice(2011)]{Kalbfleisch2011}
J.~D. Kalbfleisch and R.~L. Prentice.
\newblock \emph{The statistical analysis of failure time data}, volume 360.
\newblock John Wiley \& Sons, 2011.

\bibitem[Li and Lu(2015)]{Li2015}
G.~Li and X.~Lu.
\newblock A bayesian approach for instrumental variable analysis with censored
  time-to-event outcome.
\newblock \emph{Statistics in Medicine}, 34\penalty0 (4):\penalty0 664--684,
  2015.

\bibitem[Li et~al.(2015)Li, Fine, and Brookhart]{LiFine2015}
J.~Li, J.~Fine, and A.~Brookhart.
\newblock Instrumental variable additive hazards models.
\newblock \emph{Biometrics}, 71\penalty0 (1):\penalty0 122--130, 2015.

\bibitem[Li and Gray(2016)]{Li2016}
S.~Li and R.~J. Gray.
\newblock Estimating treatment effect in a proportional hazards model in
  randomized clinical trials with all-or-nothing compliance.
\newblock \emph{Biometrics}, 3\penalty0 (72):\penalty0 742--750, 2016.

\bibitem[Loeys and Goetghebeur(2003)]{Loeys2003}
T.~Loeys and E.~Goetghebeur.
\newblock A causal proportional hazards estimator for the effect of treatment
  actually received in a randomized trial with all-or-nothing compliance.
\newblock \emph{Biometrics}, 59\penalty0 (1):\penalty0 100--105, 2003.

\bibitem[MacKenzie et~al.(2016)MacKenzie, L{\o}berg, and
  O’Malley]{Mackenzie2016}
T.~A. MacKenzie, M.~L{\o}berg, and A.~J. O’Malley.
\newblock Patient centered hazard ratio estimation using principal
  stratification weights: application to the norccap randomized trial of
  colorectal cancer screening.
\newblock \emph{Observational Studies}, 2:\penalty0 29--50, 2016.

\bibitem[Martinussen et~al.(2016)Martinussen, Vansteelandt, Tchetgen, and
  Zucker]{Martinussen2016}
T.~Martinussen, S.~Vansteelandt, E.~Tchetgen, and D.~M. Zucker.
\newblock Instrumental variables estimation of exposure effects on a
  time-to-event response using structural cumulative survival models.
\newblock \emph{arXiv preprint arXiv:1608.00818}, 2016.

\bibitem[Nie et~al.(2011)Nie, Cheng, and Small]{Nie2011}
H.~Nie, J.~Cheng, and D.~S. Small.
\newblock Inference for the effect of treatment on survival probability in
  randomized trials with noncompliance and administrative censoring.
\newblock \emph{Biometrics}, 67\penalty0 (4):\penalty0 1397--1405, 2011.

\bibitem[Prorok et~al.(2000)Prorok, Andriole, Bresalier, Buys, Chia, Crawford,
  and et~al.]{prorok}
P.~C. Prorok, G.~L. Andriole, R.~S. Bresalier, S.~S. Buys, D.~Chia, E.~D.
  Crawford, and et~al.
\newblock Design of the prostate, lung, colorectal and ovarian (plco) cancer
  screening trial.
\newblock \emph{Controlled Clinical Trials}, 21\penalty0 (6, Supplement
  1):\penalty0 273S -- 309S, 2000.
\newblock ISSN 0197-2456.
\newblock \doi{https://doi.org/10.1016/S0197-2456(00)00098-2}.
\newblock URL
  \url{http://www.sciencedirect.com/science/article/pii/S0197245600000982}.

\bibitem[{R Core Team}(2017)]{rbase}
{R Core Team}.
\newblock \emph{R: A Language and Environment for Statistical Computing}.
\newblock R Foundation for Statistical Computing, Vienna, Austria, 2017.
\newblock URL \url{https://www.R-project.org/}.

\bibitem[Robins and Tsiatis(1991)]{Robins1991}
J.~M. Robins and A.~A. Tsiatis.
\newblock Correcting for non-compliance in randomized trials using rank
  preserving structural failure time models.
\newblock \emph{Communications in statistics-Theory and Methods}, 20\penalty0
  (8):\penalty0 2609--2631, 1991.

\bibitem[Rousseeuw and Croux(1993)]{Rousseeuw1993}
P.~J. Rousseeuw and C.~Croux.
\newblock Alternatives to the median absolute deviation.
\newblock \emph{Journal of the American Statistical association}, 88\penalty0
  (424):\penalty0 1273--1283, 1993.

\bibitem[Siegel et~al.()Siegel, Miller, and Jemal]{acs}
R.~L. Siegel, K.~D. Miller, and A.~Jemal.
\newblock Cancer statistics, 2018.
\newblock \emph{CA: A Cancer Journal for Clinicians}, 68\penalty0 (1):\penalty0
  7--30.
\newblock \doi{10.3322/caac.21442}.
\newblock URL \url{https://onlinelibrary.wiley.com/doi/abs/10.3322/caac.21442}.

\bibitem[Tchetgen et~al.(2015)Tchetgen, Walter, Vansteelandt, Martinussen, and
  Glymour]{Tchetgen2015}
E.~J.~T. Tchetgen, S.~Walter, S.~Vansteelandt, T.~Martinussen, and M.~Glymour.
\newblock Instrumental variable estimation in a survival context.
\newblock \emph{Epidemiology (Cambridge, Mass.)}, 26\penalty0 (3):\penalty0
  402, 2015.

\bibitem[Therneau(2015)]{rsurvival}
T.~M. Therneau.
\newblock \emph{A Package for Survival Analysis in S}, 2015.
\newblock URL \url{https://CRAN.R-project.org/package=survival}.
\newblock version 2.38.

\bibitem[van~der Vaart and Wellner(1996)]{Van1996}
A.~W. van~der Vaart and J.~A. Wellner.
\newblock \emph{Weak Convergence and Empirical Processes: With Applications to
  Statistics}.
\newblock Springer Science \& Business Media, 1996.

\bibitem[Yu et~al.(2015)Yu, Chen, Sobel, and Ying]{Yu2015}
W.~Yu, K.~Chen, M.~E. Sobel, and Z.~Ying.
\newblock Semiparametric transformation models for causal inference in
  time-to-event studies with all-or-nothing compliance.
\newblock \emph{Journal of the Royal Statistical Society: Series B (Statistical
  Methodology)}, 77\penalty0 (2):\penalty0 397--415, 2015.

\bibitem[Zeng and Lin(2007)]{Zeng2007}
D.~Zeng and D.~Y. Lin.
\newblock Maximum likelihood estimation in semiparametric regression models
  with censored data (with discussion).
\newblock \emph{Journal of the Royal Statistical Society, Series B},
  69:\penalty0 507--564, 2007.

\end{thebibliography}
	
	\label{lastpage}
	
	\clearpage
	\appendix
	
	\section{Justifications for $\kappa_v$}
	
	Recall $\bU = (W, \delta, D, \xb)$. It is easy to see that
	\begin{eqnarray*}
		&&E(D(1-V)|\bU) = P(D(1-V)=1|\bU)
		= P(D_1 = D_0 = 1|\bU) P(V=0|D_1 = D_0 = 1, \bU) \\
		&&\hskip 0.in= P(D_1 = D_0 = 1|\bU) P(V=0|D_1 = D_0 = 1, W_1=min(T_1,C), \delta_1=I(T_1 \leq C), \bX) \\
		&&\hskip 0.1in = P(D_1 = D_0 = 1|\bU) P(V=0|\bX).
	\end{eqnarray*}
	The last equality uses the aforementioned assumption that censoring is independent of the instrumental variable $V$ conditional on $\xb$ and the assumption of joint independence of $(D_1, D_0, T_1, T_0)$ and $V$ conditional on $\bX$.
	
	Similarly, $E( (1-D)V | \bU ) = P(D_1 = D_0 = 0 | U) P(V=1|\bX)$. It then follows that
	
	\begin{eqnarray*}
		\kappa_v & =& E\left\{ 1 - \frac{D(1-V)}{P(V=0|\bX)} - \frac{(1-D)V}{P(V=1|\bX)} | \bU \right\} \\
		&&= 1 - P(D_1 = D_0 = 1|\bU) - P(D_1 = D_0 = 0 | \bU)
		= P(D_1 > D_0 | \bU).
	\end{eqnarray*}
	
	The result above indicates that $\kappa_v$ is always nonnegative. The justification for using the projected weight $E(\kappa|\bU)$ follows from the arguments in \cite{Abadie2002}

	\section{Proofs of Theorem \ref{thm1} and Theorem \ref{thm2}}
	{\em Proof of Theorem 1}: Define $\phi(\balpha, \bO)\equiv 1-\frac{D(1-V)}{1-\psi(\balpha, \bX)}-\frac{(1-D)V}{\psi(\balpha, \bX)}$, and
	$\bQ_n(\balpha, \bb)=n^{-1/2}\bU_{n, \phi(\balpha, \bO)}(\bb)$. Then $\bQ_n(\balpha_0, \bb)=n^{-1/2}\bU_{n, \kappa}(\bb)$ and $\bQ_n(\hat\balpha, \bb)=n^{-1/2}\bU_{n, \hat\kappa}(\bb)$. Under conditions (C1)-(C3), we have
	$
	\|{\partial  \bQ_n(\balpha, \bb)}/{\partial\balpha}\|
	$
	is bounded in a neighborhood of $\balpha=\balpha_0$.  Given $\hat\balpha$ is a consistent estimator of $\ab_0$ (i.e condition (C5)), applying Taylor expansion to $\bQ_n(\hat\balpha, \bb)$ around $\balpha=\ab_0$ implies that
	\begin{eqnarray}
	\label{consistency_1}
	\sup_{\bb\in\bcalB} \|n^{-1/2}\{\bU_{n, \hat\kappa}(\bb)-\bU_{n, \kappa}(\bb)\}\|\rightarrow_{a.s.}0.
	\end{eqnarray}
	
	By the Glivenko-Cantelli Theorem \citep{Van1996}, we can show under conditions (C1)-(C2) that
	$$
	\sup_{\bb\in\bcalB, t} \|n^{-1} S_{n,\kappa}^{(j)}(\bb, t)-s_c^{(j)}(\bb, t)\|\rightarrow_{a.s.} 0,\ \ j=0, 1.
	$$
	Given condition (C3), this implies $\sup_{\bb\in\bcalB, t}  \|E_{n, \kappa}(\bb, t)-e_c(\bb, t)\|\rightarrow_{a.s.}0$.
	Then,
	\begin{eqnarray}
	&&\sup_{\bb\in\bcalB}\|\bU_{n,\kappa}(\bb)-\bar\bU_{n, \kappa}(\bb)\|\leq n^{-1/2}\sum_{i=1}^n\int_0^\infty \|E_{n, \kappa}(\bb, s)-e_c(\bb, s)\|dM_i(s)\nonumber
	\\
	&&\hskip 0.1in \leq \sup_{\bb\in\bcalB, t}  \|E_{n, \kappa}(\bb, t)-e_c(\bb, t)\|\cdot\{n^{-1/2}\sum_{i=1}^n \int_0^\infty dM_i(s)\}=o(1),\ a.s.
	\label{consistency_2}
	\end{eqnarray}
	By the results in \cite{Abadie2003} and an application of the Glivenko-Cantelli Theorem \citep{Van1996}, 
	we get
	\begin{equation}
	\sup_{\bb\in\bcalB} \|\bar\bU_{n, \kappa}(\bb)-\bmu_c(\bb)\|=o(1),\ a.s.
	\label{consistency_3}
	\end{equation}
	It follows from \eqref{consistency_1},\eqref{consistency_2}, and \eqref{consistency_3} that
	\begin{equation}
	\sup_{\bb\in\bcalB}\|n^{-1/2}\bU_{n,\hat\kappa}(\bb)-\bmu_c(\bb)\|=o(1),\ a.s.
	\label{consistency_4}
	\end{equation}
	
	By condition (C4), $\bmu_c(\bb)$ is a concave function with a unique maximizer $\bb=\bb_0$. Suppose $\hat\bb$ does not converge to $\bb_0$, a.s. Then $P({\cal E})>0$, where ${\cal E}$=\{$\exists$ a subsequence $n_k$ such that $\hat\bb_{n_k}\rightarrow \bb^*\neq \bb_0$\}. By the definition of $\hat\bb$, we have
	$n^{-1/2}\bU_{n,\hat\kappa}(\hat\bb_{n_k})\geq n^{-1/2}\bU_{n,\hat\kappa}(\bb_0)$ in ${\cal E}$, which implies $\bmu_c(\bb^*)\geq \bmu_c(\bb_0)$ given \eqref{consistency_4}. This contradicts the fact that $\bb_0$ is the unique maximizer of $\bmu_c(\bb)$. Therefore, we have $\hat\bb\rightarrow_{a.s.}\bb_0$.
	
	\vspace{1in}
	{\em Proof of Theorem 2}: Define $\bA_i(\bb)=\int_0^\infty \kappa_i \{\bZ_i-E_{n,\kappa}(\bb, s)\}dN_i(s)$, $\hat\bA_i(\bb)=\int_0^\infty \hat\kappa_i \{\bZ_i-E_{n, \hat\kappa}(\bb, s)\}dN_i(s)$. Then
	\begin{equation}\label{wc_1}
	0=\bU_{n,\hat\kappa}(\hat\bb)=\bU_{n, \kappa}(\hat\bb)+n^{-1/2}\sum_{i=1}^n \{\hat\bA_i(\hat\bb)-\bA_i(\hat\bb)\}.
	\end{equation}
	Given the consistency of $\hat\bb$, the Taylor expansion of \( \ubm_{n, \ka}(\bb) \) around $\bb=\bn$ gives
	\begin{equation} \label{wc_2}
	\ubm_{n, \ka}(\bh) \approx \ubm_{n, \ka}(\bn) - \bm{\varphi}_n(\bn)\sqrt{n}(\bh - \bn)+o(1),
	\end{equation}
	where
	$
	\bm{\varphi}_{n}(\bb) =\frac{1}{n} \sum_{i=1}^{n} \int_0^{\infty} \kappa_{i}\bV_{n, \kappa}(\bb, s) dN_{i}(s)$, and $\approx$ means the difference is $o(1),\ a.s.$
	
	On the other hand, we can write
	\begin{eqnarray}
	&&
	n^{-1/2}\sum_{i=1}^n \{\hat\bA_i(\bb)-\bA_i(\bb)\}= n^{-1/2}\sum_{i=1}^n (\hat\kappa_i-\kappa_i)\{\bZ_i-\bE_{n,\kappa}(\bb, s)\}dM_i(s)\nonumber\\
	&&\hskip 1in -n^{-1/2}\sum_{i=1}^n \hat\kappa_i \{\bE_{n, \hat\kappa}(\bb, s)-\bE_{n, \kappa}(\bb, s)\} dM_i(s).\nonumber
	\end{eqnarray}
	Define $\bD_\phi(\balpha, \bO)=\partial\phi(\balpha, \bO)/\partial\balpha^T$. Note that  $\bD_\phi(\balpha_0, \bO)$ is bounded under conditions (C2).  This implies $\sup_i |\hat\kappa_i-\kappa_i|=o(1),\ a.s.$ given condition (C5). By the Taylor expansion of $\phi(\balpha, \bO_i)$ around $\balpha=\balpha_0$ and condition (C6), we have
	\begin{eqnarray}
	n^{1/2}(\hat\kappa_i-\kappa_i)=\phi(\hat\balpha, \bO_i)-\phi(\balpha_0, \bO_i)\approx \bD_\phi(\balpha_0, \bO_i)n^{1/2}(\hat\balpha-\balpha_0)\nonumber\\
	=n^{-1/2} \sum_{j=1}^n \bD_\phi(\balpha_0, \bO_i) \bI_{\balpha}(\balpha_0, \bO_j)\equiv n^{-1/2}\sum_{j=1}^n \bI_\kappa(\balpha_0, O_j, O_i)
	\label{wc_3}
	\end{eqnarray}
	Given these results,  we can further approximate $n^{-1/2}\sum_{i=1}^n \{\hat\bA_i(\bb)-\bA_i(\bb)\}$ as follows.
	
	First, using the fact that $\sup_i |\hat\kappa_i-\kappa_i|=o(1),\ a.s.$ and applying the Glivenko-Cantelli Theorem to $\bE_{n, \kappa}$ and $\bE_{n, \hat\kappa}$, we get
	\begin{eqnarray}
	&&
	n^{-1/2}\sum_{i=1}^n \{\hat\bA_i(\bb)-\bA_i(\bb)\}\approx n^{-1/2}\sum_{i=1}^n (\hat\kappa_i-\kappa_i)\{\bZ_i-\bar\be(\balpha_0, \bb, s)\}dM_i(s)\nonumber\\
	&&\hskip 1in -n^{-1/2}\sum_{i=1}^n \kappa_i \{\bar\be(\hat\balpha, \bb, s)-\bar\be(\balpha_0, \bb, s)\} dM_i(s)\equiv (I)-(II)
	\label{wc_4}
	\end{eqnarray}
	where $\bar \be(\balpha, \bb, t)=\bar s^{(1)}(\balpha, \bb, t)/\bar \bs^{(0)}(\balpha, \bb, t)$ and $\bar\bs^{(j)}(\balpha, \bb, t)=E[\phi(\balpha, \bO)Y(t)\bZ^{\otimes j}
	$ $\exp\{\bb^T \bZ\}|D_1>D_0]$, $j=0, 1, 2$.
	
	Secondly, plugging in \eqref{wc_3} into (I), coupled with standard manipulations, leads to
	\begin{eqnarray}
	&&(I)=n^{-1/2}\sum_{i=1}^n \left(n^{-1}\sum_{j=1}^n \left[\bI_{\kappa}(\balpha_0, \bO_i, \bO_j)\int_0^\infty \{\bZ_j-\be_c(\bb, s)\}dM_j(s)\right]\right)\nonumber\\
	&&\hskip 0.2in \approx n^{-1/2}\sum_{i=1}^n \bI_{\bA,(I)}(\balpha_0, \bb, \bO_i),
	\label{wc_5}
	\end{eqnarray}
	where
	$
	\bI_{\bA,(I)}(\balpha_0, \bb, \bO_i)=E_{\bO} \left[\bI_{\kappa}(\balpha_0, \bO_i, \bO)\int_0^\infty \{\bZ-\be_c(\bb, s)\}dM(s)\right]
	$ and $E_{\bO}$ stands for expectation with respect to $\bO$.
	
	Thirdly, assessing $\bar\be(\hat\balpha, \bb, s)-\bar\be(\balpha_0, \bb, s)$ through the Taylor expansion and using condition (C6), we derive that
	\begin{eqnarray}
	&&(II)\approx n^{-1/2}\sum_{i=1}^n \left\{
	n^{-1}\sum_{j=1}^n \int_0^\infty\phi(\balpha_0, \bO_j)\bD_{\bar\be}(\balpha_0, \bb, s)dM_j(s)
	\right\}
	\bI_{\balpha}(\balpha_0, \bO_i)\nonumber\\
	&&\approx n^{-1/2} \sum_{i=1}^n \bI_{\bA, (II)}(\balpha_0, \bb, \bO_i)
	\label{wc_6}
	\end{eqnarray}
	where $\bD_{\bar\be}(\balpha, \bb, t)=\partial\bar\be(\balpha, \bb, t)/\partial\balpha$, and
	$
	\bI_{\bA, (II)}(\balpha_0, \bb, \bO_i)=E_{\bO}\{\int_0^\infty\phi(\balpha_0, \bO)\bD_{\bar\be}(\balpha_0, \bb, s)$ $dM(s)\}$ $\cdot\bI_{\balpha}(\balpha_0, \bO_i).
	$
	
	Define $\bI_{\bA}(\balpha, \bb, \bO_i)=\bI_{\bA, (I)}(\balpha, \bb, \bO_i)-\bI_{\bA, (II)}(\balpha, \bb, \bO_i)$ and $\ba_i(\bb)=\int_0^\infty \kappa_i\left\{\bZ_i-\be_c(\bb, s)\right\} $ $dM_i(s)$. It follows from \eqref{wc_1}, \eqref{wc_2}, \eqref{wc_4}, \eqref{wc_5}, and \eqref{wc_6} that
	\begin{eqnarray*}
		&&n^{1/2}(\hat\bb-\bb_0)\approx \{\varphi_n(\bb_0)\}^{-1} \{\bar\bU_{n, \kappa}(\bb_0)+n^{-1/2}\sum_{i=1}^n \bI_{\bA}(\balpha_0, \bb_0, \bO_i)\}\\
		&&\hskip 0.1in
		=n^{-1/2}\sum_{i=1}^n \{\varphi_n(\bb_0)\}^{-1} \{\ba_i(\bb_0)+\bI_{\bA}(\balpha_0, \bb_0, \bO_i)\}
	\end{eqnarray*}
	By the Central Limit Theory, we have
	$$
	n^{1/2}(\hat\bb-\bb_0)\rightarrow_d N(0, \Omega),
	$$
	where
	\begin{equation}
	\label{omega_matrix}
	\Omega=E([\{\varphi_n(\bb_0)\}^{-1} \{\ba_i(\bb_0)+\bI_{\bA}(\balpha_0, \bb_0, \bO)\}]^{\otimes 2}).
	\end{equation}

	\section{Supplemental Figures and Tables}
	
	Below we provide two figures showing the objective function and estimating function surfaces for the three proposed estimators. These were selected to demonstrate the numerical issues present in the approach and why convergence sometimes fails, and the benefits of the modified and truncated weight $\kh_{v, tr}$. The figures also demonstrate why we prefer to utilize the objective function over the estimating equation.
	
	\begin{figure*}
		\centering
		
		\includegraphics[width=.4\textwidth]{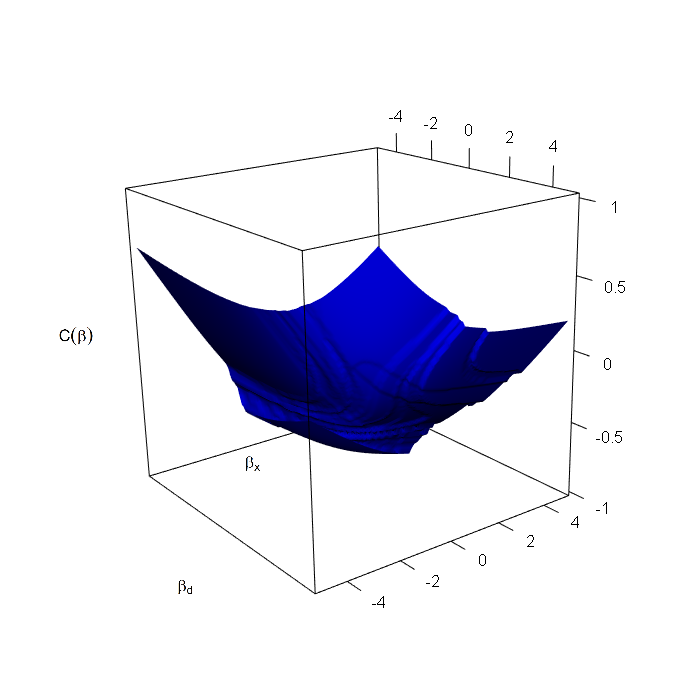}\quad
		\includegraphics[width=.4\textwidth]{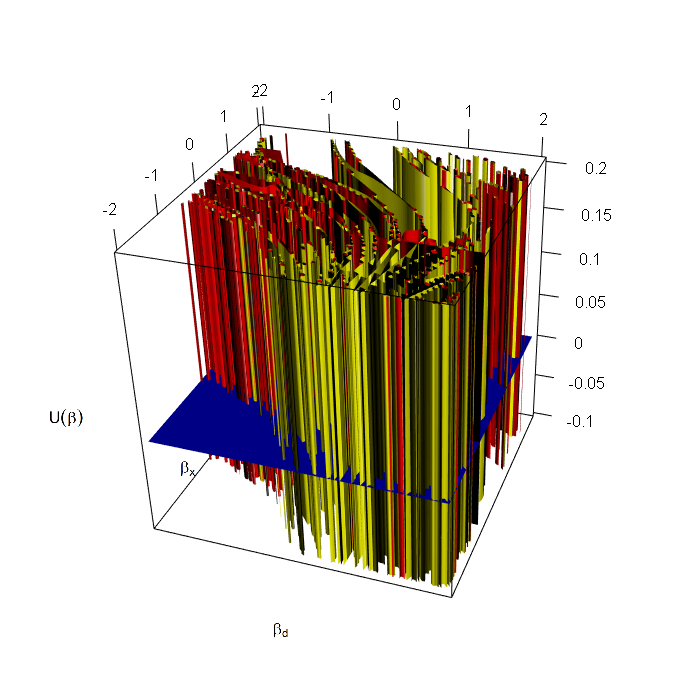}

		\medskip

		\includegraphics[width=.4\textwidth]{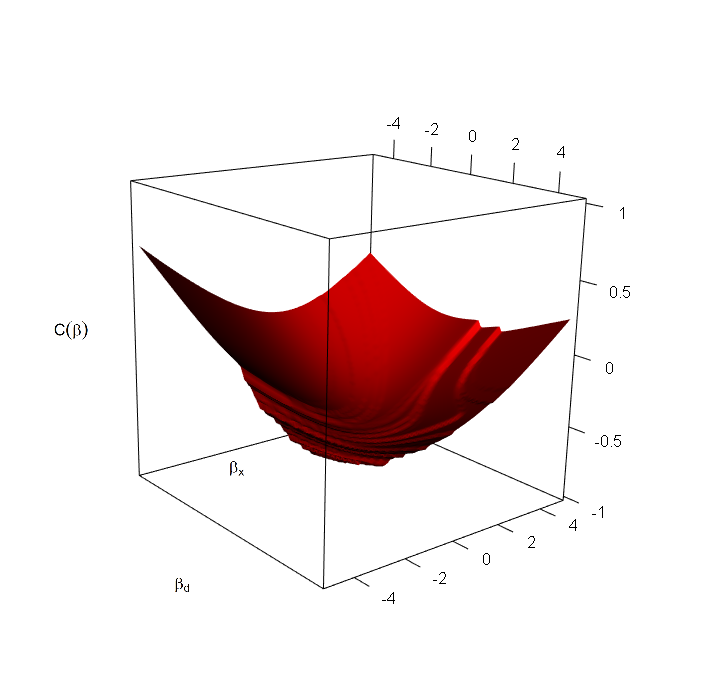}\quad
		\includegraphics[width=.4\textwidth]{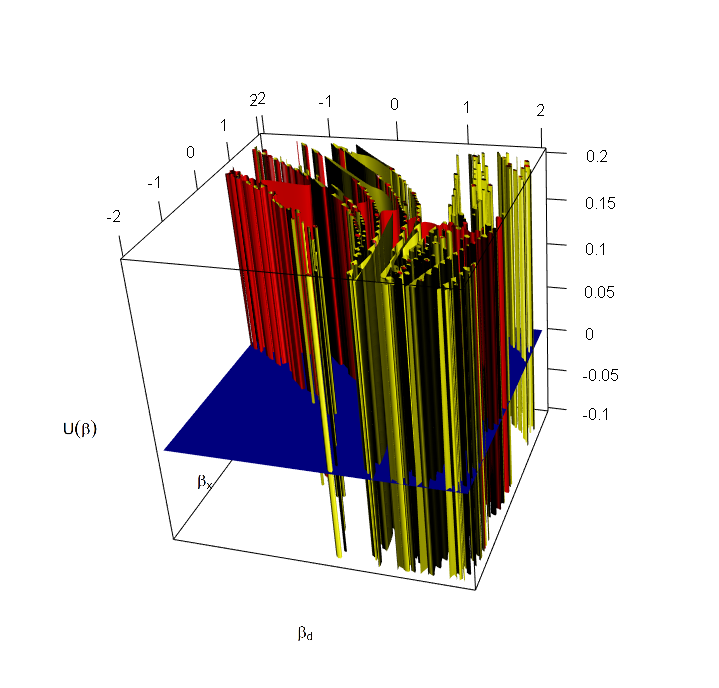}

		\medskip

		\includegraphics[width=.4\textwidth]{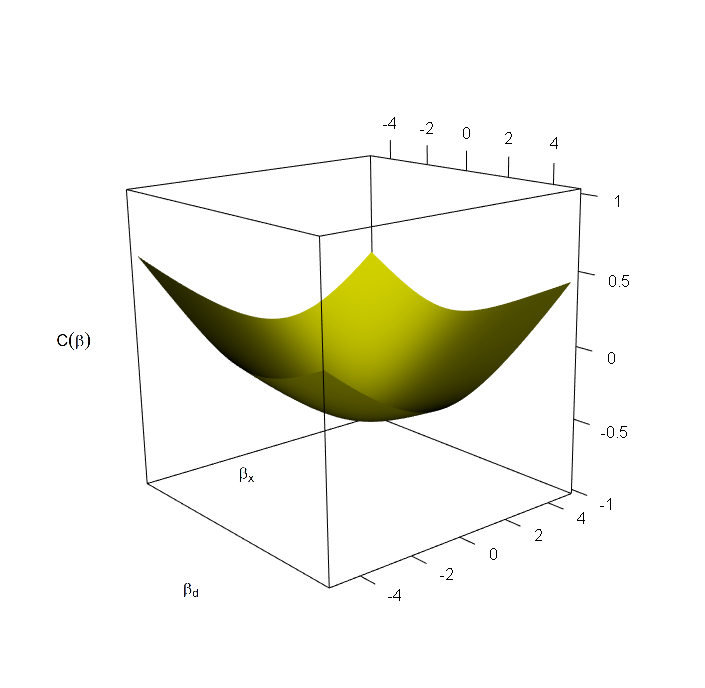}\quad
		\includegraphics[width=.4\textwidth]{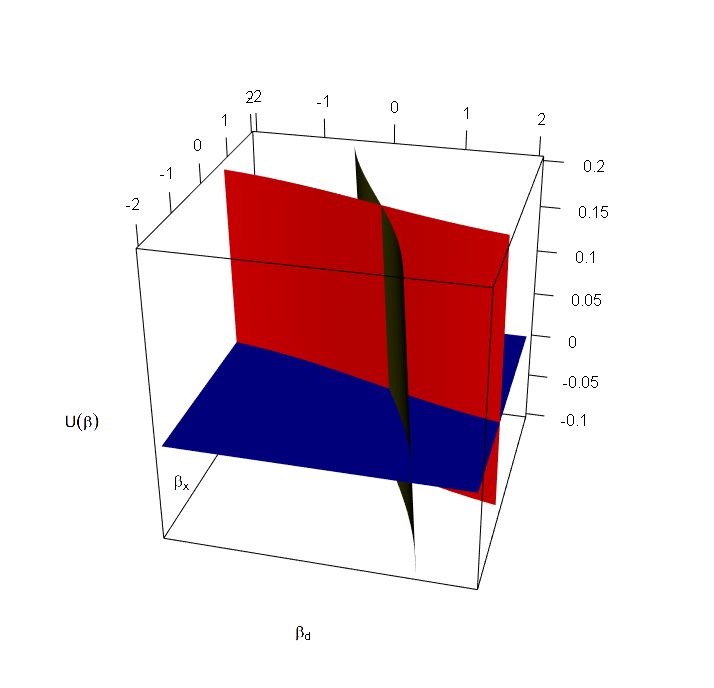}		
		
		\medskip
		
		\begin{minipage}[t]{.8\textwidth}
			\caption{Non-converged dataset from simulation 2 case 1. The left column shows the objective function for $\kappa$, $\kappa_v$, $\kappa_{v,tr}$, respectively. The right column shows the estimating function plots for methods, $\kappa$, $\kappa_v$, $\kappa_{v,tr}$, respectively}
			\label{fig:bad}
		\end{minipage}
		
	\end{figure*}
	
	\begin{figure*}
		\centering
		
		\includegraphics[width=.4\textwidth]{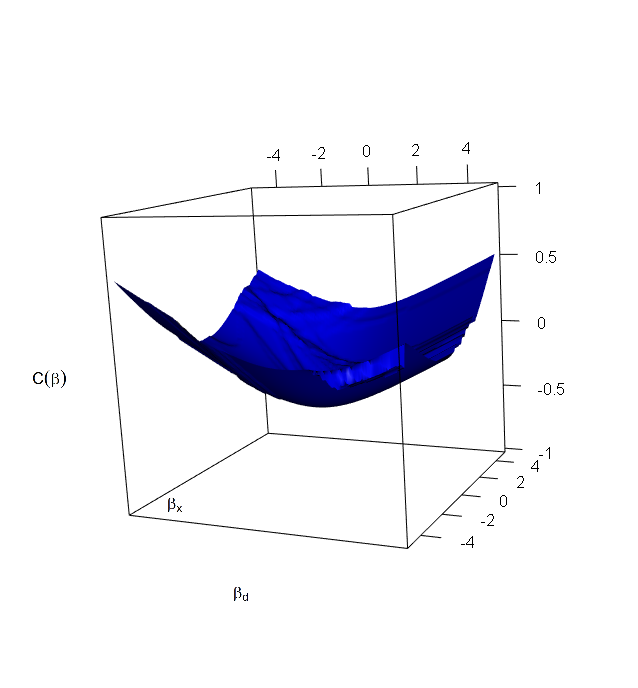}\quad
		\includegraphics[width=.4\textwidth]{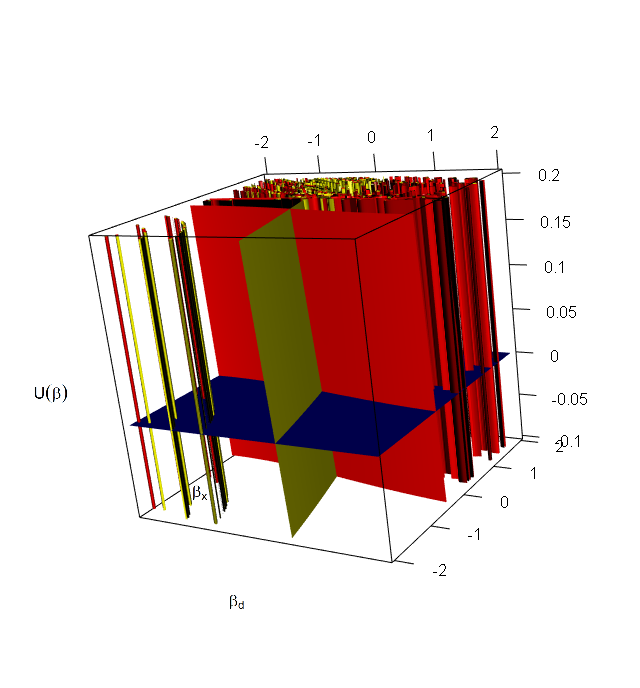}
		
		\medskip
		
		\includegraphics[width=.4\textwidth]{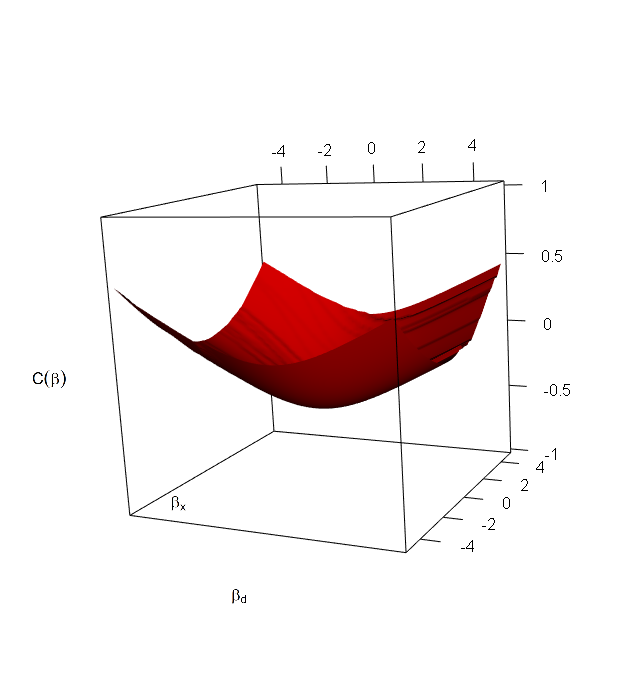}\quad
		\includegraphics[width=.4\textwidth]{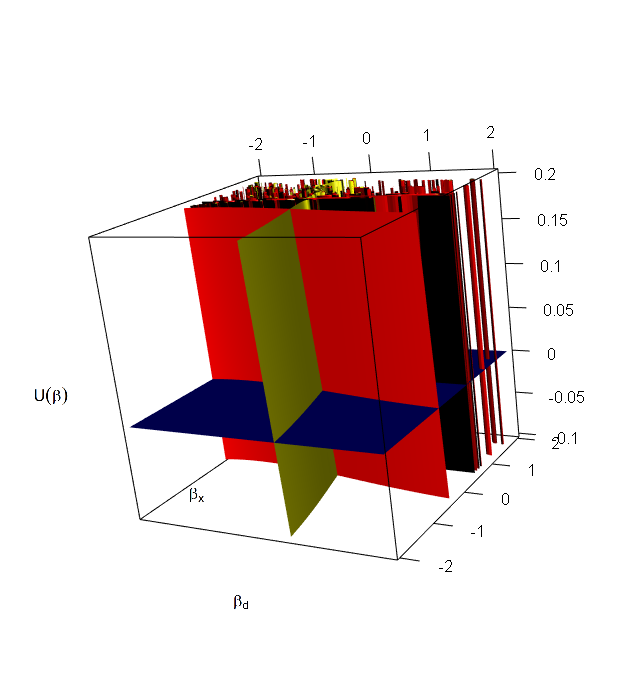}
		
		\medskip
		
		\includegraphics[width=.4\textwidth]{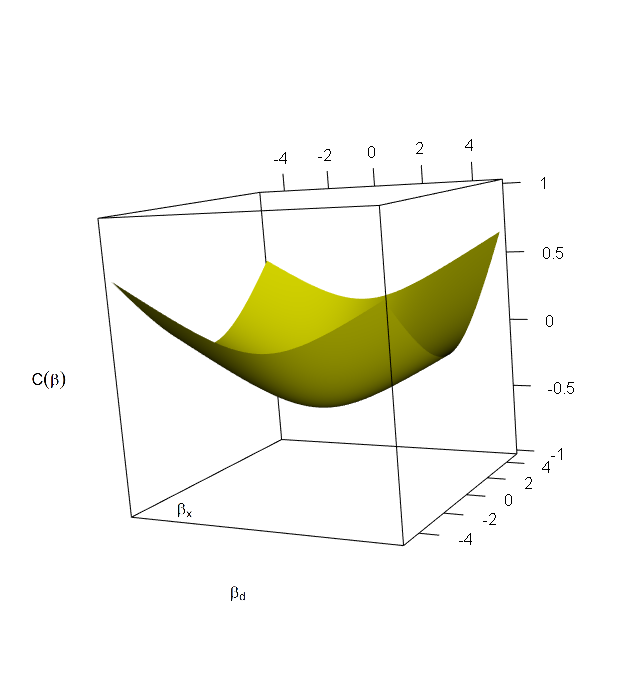}\quad
		\includegraphics[width=.4\textwidth]{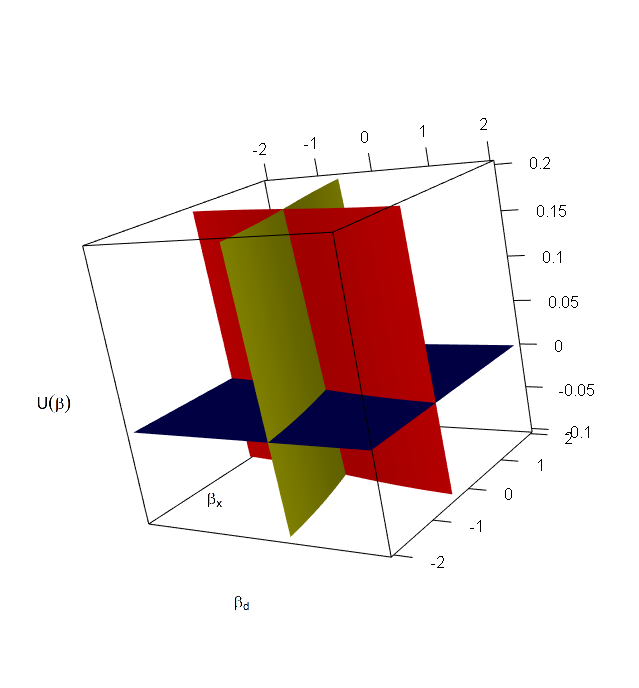}				
		\medskip
		
		\begin{minipage}[t]{.8\textwidth}
			\caption{Dataset from scenario 2 case 1 with converged estimate. The left column shows the objective function for $\kappa$, $\kappa_v$, $\kappa_{v,tr}$, respectively. The right column shows the estimating function plots for $\kappa$, $\kappa_v$, $\kappa_{v,tr}$, respectively}
			\label{fig:ok}
		\end{minipage}
		
	\end{figure*}

\end{document}